\documentclass[journal=jcisd8,manuscript=article]{achemso}

\usepackage{algorithm}
\usepackage{algpseudocode}
\usepackage{easyReview}
\usepackage{todonotes}
\usepackage{enumitem}
\usepackage{verbatim}
\usepackage{graphicx}
\usepackage{dcolumn}
\usepackage{rotating}
\usepackage{bm}
\usepackage{mathtools}

\usepackage{float}
\usepackage{graphicx}
\usepackage{subcaption}
\usepackage{hyperref}
\usepackage{multirow}
\usepackage{hhline}                         
\usepackage{amssymb, amsfonts}
\usepackage{booktabs}
\usepackage{cancel}
\usepackage{mathtools}                                
\usepackage{listings}
\usepackage{threeparttable}
\usepackage{color}
\usepackage{comment}
\usepackage{microtype}
\usepackage{kantlipsum}
\SectionNumbersOn

\newcommand\asmap{\ensuremath{\Sigma}}

\newcolumntype{P}[1]{>{\centering\arraybackslash}p{#1}}

\newcommand{\quotes}[1]{``#1''}

\author{Marco Giulini}
\affiliation{Physics Department, University of Trento, via Sommarive, 14 I-38123 Trento, Italy}
\alsoaffiliation{INFN-TIFPA, Trento Institute for Fundamental Physics and Applications, I-38123 Trento, Italy}
\altaffiliation{Present address: Bijvoet Centre for Biomolecular Research, Faculty of Science - Chemistry, Utrecht University, Padualaan 8, 3584, Utrecht, CH, The Netherlands}
\author{Raffaele Fiorentini}
\affiliation{Physics Department, University of Trento, via Sommarive, 14 I-38123 Trento, Italy}
\alsoaffiliation{INFN-TIFPA, Trento Institute for Fundamental Physics and Applications, I-38123 Trento, Italy}
\author{Luca Tubiana}
\affiliation{Physics Department, University of Trento, via Sommarive, 14 I-38123 Trento, Italy}
\alsoaffiliation{INFN-TIFPA, Trento Institute for Fundamental Physics and Applications, I-38123 Trento, Italy}
\author{Raffaello Potestio}
\affiliation{Physics Department, University of Trento, via Sommarive, 14 I-38123 Trento, Italy}
\alsoaffiliation{INFN-TIFPA, Trento Institute for Fundamental Physics and Applications, I-38123 Trento, Italy}
\email{raffaello.potestio@unitn.it}
\author{Roberto Menichetti}
\affiliation{Physics Department, University of Trento, via Sommarive, 14 I-38123 Trento, Italy}
\alsoaffiliation{INFN-TIFPA, Trento Institute for Fundamental Physics and Applications, I-38123 Trento, Italy}
\email{roberto.menichetti@unitn.it}

\title{EXCOGITO, an extensible coarse-graining toolbox for the investigation of biomolecules\\ by means of low-resolution representations}

\begin{document}

\begin{tocentry}


\includegraphics[scale=0.123]{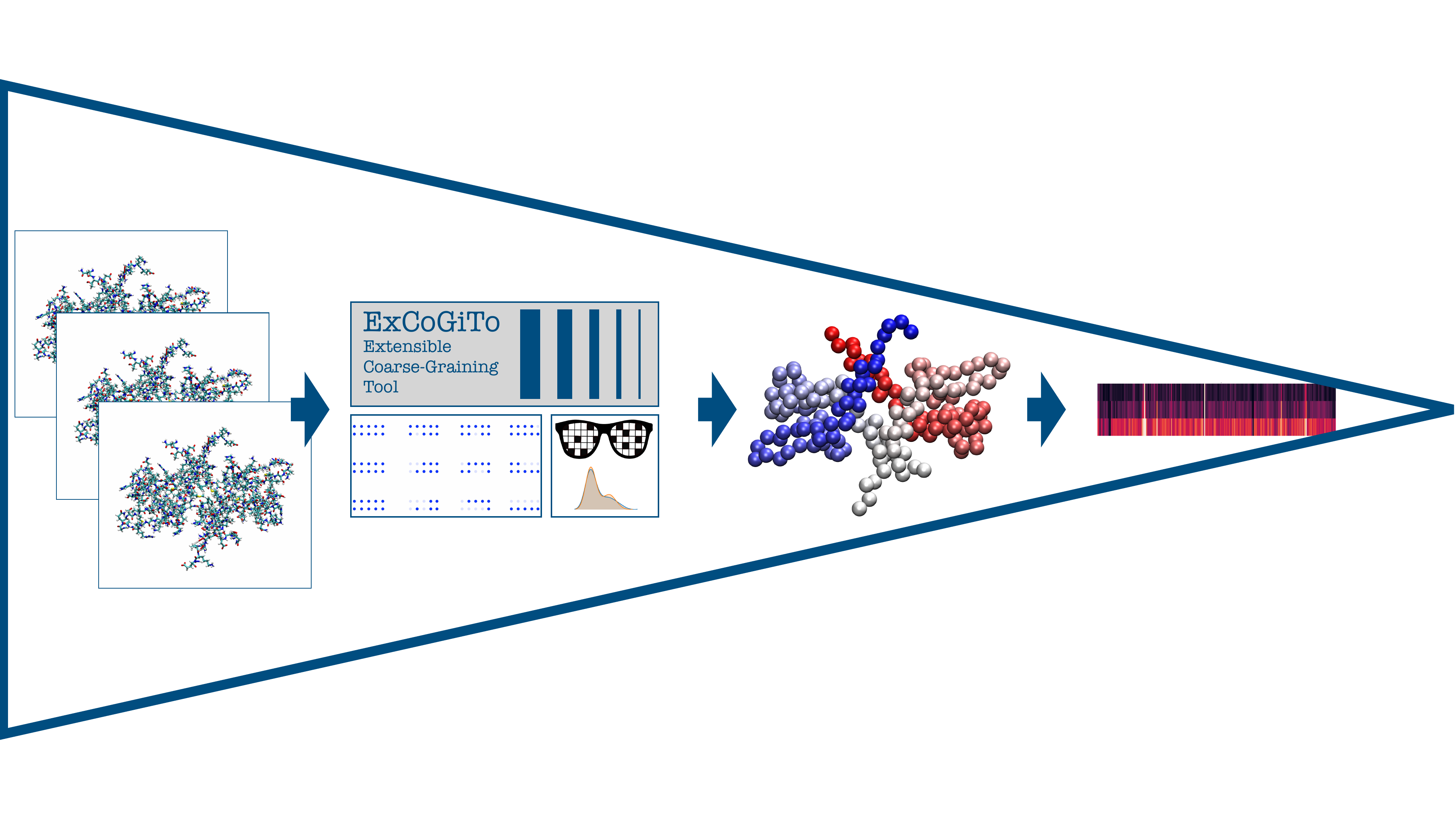}




\end{tocentry}

\date{\today}

\begin{abstract}
Bottom-up coarse-grained (CG) models proved to be essential to complement and sometimes even replace all-atom representations of soft matter systems and biological macromolecules. The development of low-resolution models takes the moves from the reduction of the degrees of freedom employed, that is, the definition of a \emph{mapping} between a system's high-resolution description and its simplified counterpart. Even in the absence of an explicit parametrisation and simulation of a CG model, the observation of the atomistic system in simpler terms can be informative: this idea is leveraged by the mapping entropy, a measure of the information loss inherent to the process of coarsening. Mapping entropy lies at the heart of the extensible coarse-graining toolbox, or EXCOGITO, developed to perform a number of operations and analyses on molecular systems pivoting around the properties of mappings. EXCOGITO can process an all-atom trajectory to compute the mapping entropy, identify the mapping that minimizes it, and establish quantitative relations between a low-resolution representation and the geometrical, structural, and energetic features of the system. Here, the software, which is available free of charge under an open-source licence, is presented and showcased to introduce potential users to its capabilities and usage.\\
Published on the \emph{J. Chem. Inf. Model.} on June 11, 2024.\\
DOI: \href{https://doi.org/10.1021/acs.jcim.4c00490}{https://doi.org/10.1021/acs.jcim.4c00490}.
\end{abstract}


\renewcommand{\baselinestretch}{1.5}
\normalsize

\clearpage


\section{Introduction} 
\label{sec:intro}

In the context of soft matter modelling, \emph{coarse-graining} (CGing) is a broad term encompassing a number of approaches, techniques, and algorithms aimed at constructing low-resolution models of molecular systems \cite{noid2013perspective, kmiecik2016coarse, giulini2021system, dhamankar2021chemically, noid2023perspective}. The objects of study can range from---structurally---simple molecules (most notably water \cite{marrink2007martini, wu2010new, hadley2012coarse}) to very complex biological machineries (proteins, DNA, lipid membranes \cite{ouldridge2010dna, marrink2023two}) up to entire cells \cite{earnest2018simulating,thornburg2022fundamental,luthey2022integrating,stevens2023molecular}. Such models are conceived so as to entail the necessary amount of information and detail to reproduce specific target properties, and enable the investigation of emergent processes and phenomena on length and time scales that would be out of reach by means of more refined descriptions \cite{potestio2014computer}, such as those employing all-atom force fields or even more accurate \emph{ab initio} approaches.

Coarse-grained modelling originates from the seminal work carried out since the 1960ies by a number of authors (most prominently Kadanoff and Wilson) in the context of the renormalisation group (RG) approach to critical phenomena \cite{wilson1971renormalization,kadanoff1990scaling,efrati2014real}; while the systems under investigation in the field of soft matter are generally far from criticality (at least in the ``standard'' sense \cite{adami1995self,mora2011biological,marsili2020importance}) and bear little if any similarity with the scale-invariant ones at the critical point, certain \emph{technical} aspects of the RG have been inherited in their study, specifically the process of \emph{mapping}.

In fact, bottom-up CG modeling \cite{noid2013perspective, jin2022bottom} requires, as a first step, that one identifies a formal map between a high-resolution description of the system and the low-resolution counterpart; such maps, which are direct descendants of Kadanoff's block-spin RG approach, are necessary prerequisites for the construction of a coarse model of a polymer, a protein, or any other molecular system. In general, a relatively small group of high-resolution constituents of the system (e.g. atoms) are lumped together in \emph{CG sites} whose properties, in particular their positions, are functions of those of the particles they represent.

Once this mapping has been defined, the subsequent step consists in the definition of the effective interactions among CG sites. In the past few decades, a number of methods \cite{noid2013perspective,brini2013systematic,saunders2013coarse,potestio2014computer,ingolfsson2014power,noid2023perspective,giulini2021system} have been devised to construct, parametrise, approximate interactions entailing the effect of the degrees of freedom that have been integrated out, and give rise to the phenomenology (or at least a behaviour close to it) one would expect from the underlying, high-resolution model.

In contrast with the intense effort invested in the development of CG \emph{force fields}, much less work has been done to investigate the properties of mappings themselves. Only recently researchers have focused on the relationships that exist between the properties of the mapping and those of the CG model that relies on it \cite{rudzinski2014investigation, wang2019coarse, yang2023slicing}; furthermore, interest is growing on the properties \emph{of the reference, high-resolution system itself} that can be learnt and rationalised in terms of a low-resolution representation.

Indeed, the process of filtering the high-resolution model of the system through the mapping can be very informative \emph{per se}. By definition, a low-resolution representation of a system entails a lower amount of information about it with respect to the full, high-resolution picture. However, it is generally the case that the large amount of detail contained in the latter hides or obscures the relevant information, that is, the salient features one needs in order to build a simple, mechanistic understanding of the system's inner workings. Hence, a mapping can be informative when the amount of information it filters in is maximised over all possible ways of discarding part of the system's structure.

In order to find those highly informative low-resolution representations a method is needed to quantify how much information is retained by them. This can be done through mapping entropy \cite{shell2008relative, foley2015impact, giulini2020information, kidder2021energetic}, which is defined as the Kullback-Leibler divergence between the (empirical) probability density of high-resolution configurations and the low-resolution counterpart obtained through the mapping. In previous works \cite{giulini2020information, holtzman2022making}, some of us have shown that those mappings that minimize the mapping entropy bear nontrivial and useful knowledge about the system and its function. This is a critical point, in that the notion of a mapping's informativeness is solely based on the conformational space explored by the system, while the information it provides can be traced back to the physical, chemical, and possibly biological properties of the object of study.

Mappings can be useful to characterise the system even in the absence of a sampling of its conformational space. Indeed, a measure of distance between mappings can be leveraged to highlight structural features of a molecule and explore the \emph{mapping space} in a quantitative manner \cite{menichetti2021journey}.

\begin{figure}
\centering
    \includegraphics[width=\columnwidth]{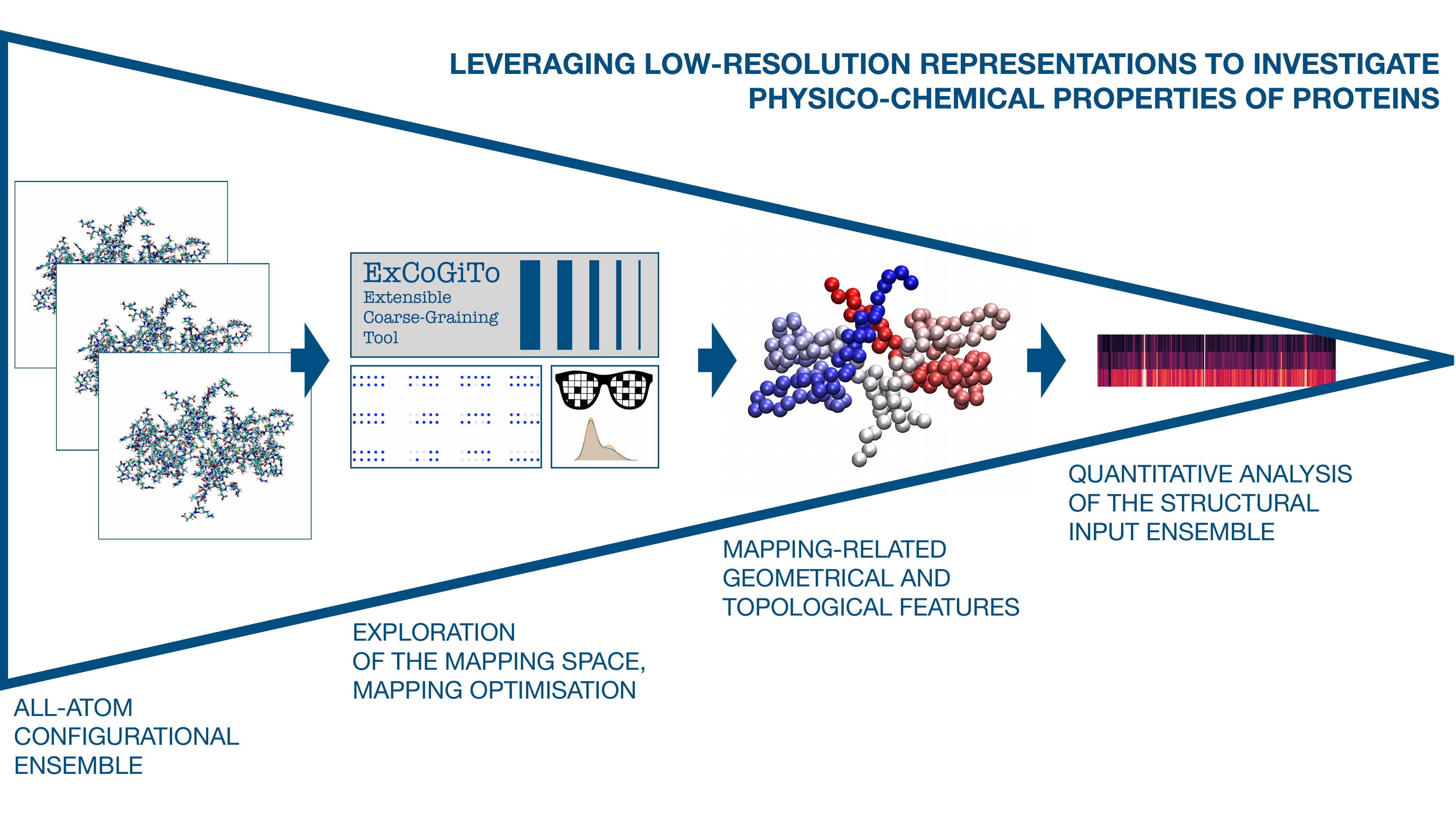}
\caption{\label{fig:simplescheme}A pictorial representation of EXCOGITO and its purpose. The software carries out various types of analyses on a sample of high-resolution, or all-atom, protein configurations. These are investigated in terms of reduced representations, or mappings, in which only a subset of the molecule's atoms are retained. This kind of analysis can provide novel and useful information on various properties of the system, its conformational variability, its energetics, and how all these properties can be related to the biological function. For further details and examples \emph{vide infra} as well as Refs. \citenum{giulini2020information,menichetti2021journey}.}
\end{figure}

Key to the fruitful usage of these concepts and methods to all applications illustrated insofar, as well as many others, is their implementation in an efficient and easy-to-use software platform. In this work, we present, describe, and showcase the extensible coarse-graining toolbox, or EXCOGITO, developed to provide users with the necessary instruments to make the most of the concept of mapping and mapping entropy. EXCOGITO, whose workflow and purpose are schematically represented in Fig. \ref{fig:simplescheme}, implements several tools that allow the investigation of complex molecular systems through various instruments all pivoting on the concept of mapping, leveraging the idea that a relation exists between the most informative low-resolution representation of a system and its physical properties.

In the following we provide the theoretical foundations of the methods implemented in the software, then proceed to describe it and how it works through its application to specific case studies. We review the previous literature about mapping entropy minimization and mapping space exploration that leverages tools implemented in EXCOGITO, and show the application of these methods to a simple yet nontrivial system, icosalanine.

EXCOGITO, written in C, is free to download, simple to use, and provides researchers with a novel and powerful instrument to investigate the properties of complex biological or artificial macromolecules.

\section{Methodology}

The aim of the EXCOGITO software is to provide users with a simple and efficient tool to investigate the features of the mapping space of a macromolecular system, most notably proteins, and explore the relation they entertain with the physical and biological properties of the latter. This approach relies on a simple yet powerful idea, namely that \emph{by losing resolution}, e.g. by looking at the system in terms of fewer atoms than the total, \emph{we gain information} about the processes that take place at a more global scale. Since a few years, this idea is put forward by various authors \cite{rudzinski2014investigation,foley2015impact,foley2020exploring,giulini2020information,menichetti2021journey,holtzman2022making,kidder2024surveying}; here we do not aim to review the results obtained insofar, nor to discuss the theoretical and practical aspects of its implementation. Rather, we illustrate the software we developed to put this idea to work.

In the following, we will briefly recall the definition and properties of the fundamental theoretical ingredient that lies at the core of EXCOGITO, namely the mapping entropy $S_{map}$ quantifying the loss of statistical information generated by coarse-graining a macromolecular system. We will then describe the $S_{map}$-based tools that are currently implemented in the software, whose applications range from computing the information loss associated with a specific CG representation to the minimization of $S_{map}$ in the space of possible low-resolution descriptions. For this latter protocol, we will further discuss how a statistical analysis performed over those representations that minimize the information loss with respect to the high-resolution reference can provide insight into the system's biological properties. Before introducing the concept of mapping entropy and the related EXCOGITO analysis tools, however, it is crucial to understand why and how an information loss arises if one blurs the description of the system of interest. We thus start summarising the basic principles underlying coarse-graining procedures.

\subsection{Information loss in coarse-graining}
\label{sec:infloss}

We consider a macromolecular system composed of $n$ constituent atoms---comprising the solute(s) as well as the solvent ones---modelled as point-like particles mutually interacting \emph{via} classical potentials. The set of Euclidean coordinates $\mathbf{r}_i$, $i=1,...,n$ of all atoms defines the high-resolution configuration $\mathbf{r}$, or \emph{microstate}, of the system to which we associate an atomistic probability distribution $p(\mathbf{r})$. We stress that $p(\mathbf{r})$ is in principle arbitrary, in the sense that no assumption needs to be made on the functional form of this probability and/or the nature of the underlying generative process. As a general but relevant example, under the hypothesis of thermal equilibrium one can assume that $p(\mathbf{r})$ is a Boltzmann probability density, i.e. the system is described by the canonical ensemble, with
\begin{equation}
\label{eq:boltzmanndist}
p(\mathbf{r})=\frac{e^{-\beta u(\mathbf{r})}}{Z}.
\end{equation}
In Eq.~\ref{eq:boltzmanndist}, $u(\mathbf{r})$ is the interaction potential among the atoms comprising non-bonded (van der Waals, electrostatic...) and bonded (bonds, angles, dihedrals...) contributions, $\beta=1/k_B T$ is the inverse temperature, and $Z$ is the configurational partition function,
\begin{equation}
\label{eq:partition}
Z = \int d\mathbf{r}\ e^{-\beta u(\mathbf{r})}.
\end{equation}

Starting from the fully-atomistic picture, low-resolution or CG representations of the system are obtained by lumping together groups of atoms into effective interaction sites, thus resulting in a reduction in the level of detail at which the macromolecule is observed. Practically, this is achieved through the introduction of a mapping operator $\mathbf{M}$ that projects a high-resolution configuration $\mathbf{r}$ of the system onto its low-resolution counterpart $\mathbf{R}=\mathbf{M}(\mathbf{r})$, the latter being defined only in terms of the coordinates $\mathbf{R}_I$, $I=1,...,N$, of the $N<n$ effective sites chosen ($N$ being often referred to as ``\emph{degree of coarse-graining}'' \cite{dhamankar2021chemically}), with
\begin{equation}
\label{eq:mappingdef}
\mathbf{R}_I=\mathbf{M}_I(\mathbf{r})=\sum_{i\in I}c_{Ii}\mathbf{r}_i, \ \ \  I=1,...,N.
\end{equation}
The linear coefficients in Eq.~\ref{eq:mappingdef} are constant, positive, and satisfy the normalization condition $\sum_{i\in I}c_{Ii}=1$ to preserve translational invariance. Furthermore, it is often assumed that different CG sites do not have atoms in common, so that if $c_{Ii}\neq0$ and atom $i$ contributes to the position of site $I$ one has $c_{Ji}=0 \ \  \forall J\neq I$.

A particular CG representation of the system is obtained for a specific choice of $N$ and of the set of coefficients $c_{Ii}$; by varying these ingredients one spans the so-called \emph{mapping space}, namely the ensemble of all possible reduced representations that can be constructed to describe the macromolecule. Importantly, in the following we will restrict our attention to \emph{decimation mappings}, in which a subset of $N<n$ atoms of the macromolecule are retained as low-resolution CG sites, while the remainder (solvent included) is neglected; this procedure is implemented through a set of selection operators $\chi_{{\bf{M}},i}$,\  $i=1,...,n$:
\begin{eqnarray}
\label{eq:mapping}
\chi_{{\bf M},i} &&=
\left\{
	\begin{array}{ll}
		1  & \mbox{if atom $i$ is retained,}\\
		0 & \mbox{if atom $i$ is not retained,}
	\end{array}
\right.\\
\label{eq:mapping_degree}
&&\;\;\;\;\;\;\;\;\sum_{i = 1}^n \chi_{{\bf M},i} = N.
\end{eqnarray}

Irrespective of the particular choice of $\mathbf{M}$, the projection performed by the mapping operator lies at the core of the information loss generated by coarse-graining. To understand why, we note that the transformation in Eq.~\ref{eq:mappingdef} is non-invertible: while each atomistic configuration is associated with a single low-resolution one, the opposite does not hold, and a given CG configuration is actually compatible with a whole \emph{pool} of possible microstates, namely all those in which the coordinates of the discarded atoms vary while keeping the positions of the retained sites fixed. For an observer who examines the system only \emph{via} the ``filtered'', projected configurations, such microstates are in all respects indistinguishable, and grouped together they constitute what is commonly referred to as a CG \emph{macrostate} $\mathbf{R}$. The probability $P(\mathbf{R})$ that the observer will associate with a specific macrostate reads
\begin{eqnarray}
\label{eq:pcoarse}
P({\bf R}) = \int d\mathbf{r}\ p({\bf r}) \delta({\bf M}({\bf r}) - {\bf R}),
\end{eqnarray}
and is thus obtained by integrating over all the high-resolution configurations ${\bf r}$ of the system that, upon the projection ${\bf M}({\bf r})$, are mapped onto macrostate ${\bf R}$, each configuration being weighted with its atomistic probability $p({\bf r})$. Apart from some general features inherited from the fundamental symmetries characterizing the high-resolution system---such as, e.g., rotational and translational invariance---the CG probability $P(\mathbf{R})$ will critically depend on the prescription employed to group together microscopic configurations to form the macrostates, that is, on the decimation mapping operator $\mathbf{M}$; it is thus easy to imagine that the choice of the CG mapping will determine how much information on the system properties will be transferred from the high- to the low-resolution representation, and that understanding why a mapping is ``better'' than another can lead to a deeper understanding of the system.

In the next section, we thus address the question of how to select mappings that are ``better'' than others, specifically starting from the problem of quantifying in an unambiguous manner the quality of a mapping.

\subsection{Mapping entropy and related EXCOGITO tools}
\label{sec:mapentrtools}

Eq.~\ref{eq:pcoarse} represents the elemental equation of coarse-graining, in that it enables---al least theoretically---to determine the low-resolution properties of a system starting from the laws that govern the statistical behavior of its microscopic constituents. Consider now, however, an attempt of \emph{reverting} this procedure, with the observer who only collects knowledge of the low-resolution distribution $P(\mathbf R)$ aiming at reconstructing the high-resolution detail of the system, namely the fully-atomistic probability distribution $p(\mathbf{r})$. As previously discussed, for each CG macrostate $\mathbf{R}$ the specific properties of the microstates that enter its composition have been lost along the projection; only provided with the cumulative probability of each macrostate and the connection between the high and low-resolution configurational ensembles $\mathbf{R}=\mathbf{M}(\mathbf{r})$, the most sensible and potentially only choice left to the observer for reconstructing the atomistic distribution should then be compatible with a \emph{maximum entropy principle}, in which all the microscopic configurations that belong to a particular CG macrostate are equally likely to occur. Accordingly, the resulting \emph{backmapped} atomistic probability distribution $\bar{p}_{r}(\mathbf{r})$ reads 
\begin{equation}
\label{eq:barp}
\bar{p}_r({\bf r}) = \frac{P({\bf M}({\bf r}))}{\Omega_1 ({\bf M}({\bf r}))},
\end{equation}
where
\begin{equation}
\label{eq:omega1}
\Omega_1({\bf R})=\int d\mathbf{r} \ \delta({\bf M}({\bf r}) - {\bf R}) 
\end{equation}
is the number of microstates ${\bf r}$ mapping onto the CG macrostate ${\bf R}$. It follows that $\bar{p}_r(\mathbf{r})$ constitutes a smeared version of the original distribution, where, in contrast to the latter, all configurations that map onto the same macrostate are endowed with the same statistical weight, this being equal to the average of the original probabilities of these microstates. Reverting the coarse-graining procedure has hence introduced a bias in the statistical properties of the backmapped high-resolution system. 

In information-theoretical approaches, if a system originally described by a probability distribution $s(\mathbf{r})$ is represented in terms of a different one $t(\mathbf{r})$, the associated loss of statistical information can be quantified \emph{via} the Kullback-Leibler (KL) divergence $D_{KL}(s || t)$\cite{kullback1951information}, with
\begin{equation}
D_{KL}(s || t)=\int d\mathbf{r} \ s(\mathbf{r}) \ln\left[\frac{s(\mathbf{r})}{t(\mathbf{r})}\right].
\end{equation}

$D_{KL}(s || t)$ can be considered a ``distance'' in probability space between the two distributions, where the quotes account for the fact that $D_{KL}$ is non-symmetric with respect to the exchange of $s(\mathbf{r})$ and $t(\mathbf{r})$. By virtue of Gibbs' inequality one has $D_{KL}(s || t)\geq 0$ for all $s(\mathbf{r}), t(\mathbf{r})$, where $D_{KL}(s || t)=0$ only if $t(\mathbf{r})=s(\mathbf{r})$. In the case of a coarse-graining procedure performed on the system through a decimation mapping, see Eq.~\ref{eq:mapping}, the KL divergence between the original and reconstructed probability distributions $p(\mathbf{r})$ and $\bar{p}_r(\mathbf{r)}$ is dubbed \emph{mapping entropy} $S_{map}$,\cite{shell2008relative,rudzinski2011coarse,foley2015impact} 
\begin{equation}
\label{eq:smap}
S_{map}({\bf M}) = k_B D_{KL}(p || \bar{p}_r) = k_B\int d\mathbf{r}\ p({\bf r}) \ln \left[ \frac{p({\bf r})}{\bar{p}_r({\bf r})} \right]=k_{B}\biggl<\ln \left[\frac{p({\bf r})}{\bar{p}_r({\bf r})}\right]\bigg>
\end{equation}
and represents a measure of the loss of information \emph{inherently generated by the structural coarsening}. In Eq.~\ref{eq:smap} the average $\langle \cdot \rangle$ is performed over the high-resolution probability distribution, and we further emphasize the dependency of $S_{map}$ on the choice of the mapping operator $\mathbf{M}$---that is, on the location and amount of retained atoms, see Eqs.~\ref{eq:mappingdef} and \ref{eq:mapping_degree}---to underline that, in general, different low-resolution representations carry a different amount of information about the system. Critically, this opens the possibility of investigating whether simplified CG representations exist that \emph{minimize} the mapping entropy, thus being capable of retaining the maximum amount of information on the statistical properties of the macromolecule despite a reduction in the level of detail employed to describe it. The identification of such \emph{maximally informative} mappings naturally passes through the possibility of calculating $S_{map}$ for a specific choice of the CG representation of the system; firstly, let us then focus on the tools currently implemented in EXCOGITO to achieve this task, which constitutes the fundamental building block for the more advanced $S_{map}$-based analysis workflows to be described in the following.

 In principle, given a CG mapping the associated $S_{map}$ can be directly evaluated through the definition in Eq.~\ref{eq:smap} provided that the all-atom probability distribution $p(\mathbf{r})$ is known, its backmapped counterpart $\bar{p}_r(\mathbf{r)}$ can be explicitly determined and the summation over the microstates $\mathbf{r}$ exhaustively performed. This is the case, for example, of the coarse-graining of simple systems characterized by a finite and low-dimensional configurational space such as discrete classical spins on a small lattice \cite{holtzman2022making}. When considering complex macromolecules such as proteins---that is, the main target of the current release of EXCOGITO---however, even in the case of a known functional form of the probability distribution, the calculation of Eq.~\ref{eq:smap} would require solving analytically intractable high-dimensional integrals over the coordinates of the constituent atoms.
 
 Two possible ways of tackling the problem are currently available in EXCOGITO; the associated tools, respectively called \emph{\textbf{measure}} and \emph{\textbf{measure\_kl}}, enable the user to approximately estimate the mapping entropy associated with the resolution reduction of a macromolecular system starting from a set of ingredients that have to be provided in input to the software. As we illustrate hereafter, these ingredients amount to the configurations of the system under examination, and \emph{either} the potential energy associated to each configuration (\emph{\textbf{measure}} tool), \emph{or} the value of its statistical weight (\emph{\textbf{measure\_kl}} tool). In the following subsections, we describe these different approaches, their conceptual frameworks, and contexts of applicability.

\noindent {\bf CUMULANT EXPANSION STRATEGY |} The first EXCOGITO $S_{map}$ calculation protocol, implemented in the \emph{\textbf{measure}} tool, relies on configurational sampling and applies to an equilibrium condition in which the high-resolution probability $p(\mathbf{r})$ of the macromolecule is given by the Boltzmann distribution, see Eq.~\ref{eq:boltzmanndist}. To illustrate the underlying method, let us first consider the case of an arbitrary $p(\mathbf{r})$---whose analytical form is known---and assume that a discrete series of atomistic configurations $\mathbf{r}_i$, $i=1,...,K\gg1$ of the system sampled from such distribution \emph{via}, e.g., Molecular Dynamics or Monte Carlo simulations is available. Given these configurations and the choice of the CG representation, $S_{map}$ could in principle be estimated as
\begin{equation}
\label{eq:smap_disc}
S_{map}({\bf M}) =  \frac{1}{K}\sum_{i=1}^K k_B\ln \left[ \frac{p({\bf r}_i)}{\bar{p}_r({\bf r}_i)} \right].
\end{equation}
Two main criticalities unfortunately arise in Eq.~\ref{eq:smap_disc}, namely that (\emph{i}) the backmapped probability $\bar{p}_r({\bf r})$ is a highly non-local function of the all-atom distribution $p(\mathbf{r})$, see Eqs.~\ref{eq:pcoarse}-\ref{eq:omega1}; and (\emph{ii}) even if $\bar{p}_r({\bf r})$ is known, the logarithm of the ratio in Eq~\ref{eq:smap_disc} is still prone to numerical instabilities. At the same time, for equilibrium systems in which $p(\mathbf{r})\propto \exp[-\beta u(\mathbf{r})]$ some of us have shown that by performing a cumulant expansion of Eq.~\ref{eq:smap} it is possible to approximate the mapping entropy as\cite{giulini2020information}
\begin{equation}
\label{eq:smap_beta}
S_{map}({\bf M}) \simeq \tilde{S}_{map}({\bf M}) = k_{B}\frac{\beta^2}{2} \int d\mathbf{R}\  P({\bf R}) \langle(u-\langle u\rangle_{{\bf R}})^2\rangle_{{\bf R}}.
\end{equation}
Eq.~\ref{eq:smap_beta} shows that $\tilde{S}_{map}$ can be calculated by first computing, for each CG macrostate $\mathbf{R}$, the variance of the \emph{atomistic} energies of all microscopic configurations that map onto it---that is to say, the term $\langle(u-\langle u\rangle_{{\bf R}})^2\rangle_{{\bf R}}$, where $\langle \cdot \rangle_{{\bf R}}$ is an equilibrium average conditioned to the macrostate. Subsequently, such variances have to be averaged over all possible CG macrostates, each one weighted with its own low-resolution probability $P(\mathbf{R})$. 
The \emph{\textbf{measure}} tool of EXCOGITO relies on the estimator in Eq.~\ref{eq:smap_beta} to evaluate the mapping entropy of a macromolecular system at equilibrium, where a set of high-resolution configuration $\mathbf{r}_i$ sampled from the Boltzmann distribution \emph{as well as} the associated atomistic energies $u(\mathbf{r}_i)$ have to be provided by the user as input to the software, together with the selected CG representation. We stress that the identification of the CG macrostates ${\bf R}$ in Eq.~\ref{eq:smap_beta} is a challenging task: in fact, it cannot be obtained by analytically marginalizing over the discarded degrees of freedom; nor would it be efficient to carry out a restrained sampling where the preserved atoms are kept fixed, since this operation would have to be repeated for a statistically significant number of CG configurations. To circumvent this limitation, \emph{\textbf{measure}} makes use of a clustering algorithm that, given the available set of high-resolution configurations $\mathbf{r}_i$, lumps them in groups based only on the atoms that are retained in the CG mapping, where such groups are then identified with the CG macrostates---a technical analysis of the different prescriptions employed by EXCOGITO to carry out this procedure being reported in Sec.~\ref{sec:confcl}. Starting from this partitioning, \emph{\textbf{measure}} computes the energy variance of each macrostate and combines together the results to calculate, via a discretized version of Eq.~\ref{eq:smap_beta}, the mapping entropy associated with a specific CG representation of the system.

\noindent {\bf KULLBACK-LEIBLER STRATEGY |} In addition to the previously described protocol that is applicable in the case of equilibrium conditions, EXCOGITO further features a different method, implemented in the \emph{\textbf{measure\_kl}} tool, to determine the mapping entropy of a macromolecule. In this second case, a discrete set of high-resolution configurations $\mathbf{r}_i$, $i=1,...,L$ of the system \emph{as well as} the associated probabilities $p(\mathbf{r}_i)$, with $\sum_{i=1}^L p(\mathbf{r}_i)=1$, have to be provided in input to the software, together with the CG representation chosen. The mapping entropy is then evaluated in \emph{\textbf{measure\_kl}} as a KL divergence over this countable state space, that is,
\begin{equation}
\label{eq:smap_kl}
\hat{S}_{map}({\bf M}) = k_B\sum_{i=1}^L\ p({\bf r}_i) \ln \left[ \frac{p({\bf r}_i)}{\bar{p}_r({\bf r}_i)} \right],
\end{equation}
where the hat superscript has been introduced to discriminate this mapping entropy estimator with the cumulant expansion $\tilde{S}_{map}$ equilibrium one reported in Eq.~\ref{eq:smap_beta}. In Eq.~\ref{eq:smap_kl}, the backmapped probabilities $\bar{p}(\mathbf{r}_i)$, $i=1,...,L$  are determined, given the selection of the atoms to be retained at the low-resolution level, by clustering all the atomistic configurations into a set of CG macrostates in analogy with the equilibrium framework, see Sec.~\ref{sec:confcl} for all technical details. Then, the weight $\bar{p}(\mathbf{r}_i)$ of the $i$-th configuration is given by the average probability of all microstates that belong to the CG cluster that contains $\mathbf{r}_i$. 

Importantly, in contrast to the $S_{map}$ and $\tilde{S}_{map}$ estimators in Eqs.~\ref{eq:smap_disc} and ~\ref{eq:smap_beta}, the defining protocol of \emph{\textbf{measure\_kl}} enables the calculation of the mapping entropy of a system also in the absence of any information on the underlying generative mechanism of the microstates, that is, on the all-atom probability $p(\mathbf{r})$. In this context, rather than frames sampled from $p(\mathbf{r})$, the $\mathbf{r}_i$ should be interpreted as \emph{representative elements} of the full configuration set, and the $p(\mathbf{r}_i)$ as frequentistic estimates of their actual probabilities.

To provide an example of when this second method can be applied and how the associated ingredients can be obtained, consider a scenario in which, although a series of $K\gg L$ all-atom configurations of the system is available, these represent samples of an unknown distribution $p(\mathbf{r})$---e.g., they are obtained from experimental measurements or emerge from MD simulations of a system in a stationary but non-equilibrium condition. In this context, neither the general estimator in Eq.~\ref{eq:smap_disc} (irrespective of its previously discussed limitations) nor the approximated $\tilde{S}_{map}$ one in Eq.~\ref{eq:smap_beta} lying at the core of \emph{\textbf{measure}} can be straightforwardly employed. However, one can perform an \emph{atomistic} clustering on this ensemble of microstates and lump them in $L$ groups based on similarity criteria. From this, each representative all-atom configuration $\mathbf{r}_i$, $i=1,...,L$ in Eq.~\ref{eq:smap_kl} can then be identified, e.g., with the centroid of a cluster, and the associated probability $p(\mathbf{r}_i)$ estimated as the fraction of configurations belonging to the cluster. Given these ingredients, the mapping entropy can finally be evaluated through Eq.~\ref{eq:smap_disc} \emph{via} an additional (this time, coarse-grained) clustering carried out on the set of $\mathbf{r}_i$ starting from a choice of the low-resolution representation of the macromolecule. We underline that \emph{\textbf{measure\_kl}} can also be employed when the functional form of the probability $p(\mathbf{r})$ is known; critically, in the case of equilibrium systems the resulting mapping entropies have been found to correlate with those obtained from the $\tilde{S}_{map}$  estimator in Eq.~\ref{eq:smap_beta},\cite{holtzman2022making} thus highlighting the robustness of this second method.

\noindent {\bf ESTIMATION OF THE QUALITY OF A CG MAPPING —} Given a series of high-resolution configurations of the system of interest endowed with either their all-atom potential energies or their ``frequentistic'' probabilities, the previously described mapping entropy calculation tools \emph{\textbf{measure}} and \emph{\textbf{measure\_kl}} enable the user of EXCOGITO to quantify the loss of statistical information experienced by a macromolecule as a consequence of a specific decimation of its microscopic degrees of freedom. One natural requirement would then be that of gauging the \emph{quality} of such CG representation based on the resulting mapping entropy; at the same time, except for its lower bound, no additional reference $S_{map}$ value can be \emph{a priori} inferred for an arbitrary system, thus hampering a straightforward interpretation of an information loss calculation performed on a single CG mapping. This problem is further compounded with the dependence of $S_{map}$ on the \emph{amount} of sites $N$ employed in the simplified description\cite{giulini2020information}---the previously introduced degree of coarse-graining, see Eqs.~\ref{eq:mappingdef} and~\ref{eq:mapping_degree}---in addition to their location throughout the molecular structure. For each analysed system and inspected number of retained sites $N$, it would thus be desirable to identify a ``characteristic scale'' of $S_{map}$ associated with  the somewhat ``typical'' reduced representations that can be constructed at that degree of coarse-graining; the quality of the proposed CG mapping can then be quantified in terms of the \emph{relative} information gain/loss that the high-resolution description such selection of sites guarantees, compared to the ones that were chosen as a reference. Critically, in the absence of any previous chemical intuition on the system, the characteristic scale should be as impartial as possible, and it is thus reasonable to deduce it from a totally unbiased exploration of the macromolecule's mapping space in which low-resolution representations with the desired number of retained sites are randomly probed.

This assessment of the typical spectrum of information loss generated by coarse-graining a macromolecular system can be performed in EXCOGITO through the \emph{\textbf{random}} or \emph{\textbf{random\_kl}} tools, which respectively rely on the previously discussed mapping entropy estimators $\tilde{S}_{map}$ and $\hat{S}_{map}$ in Eqs.~\ref{eq:smap_beta} and~\ref{eq:smap_kl}. By providing as input to the software the set of ingredients necessary for a single calculation of $S_{map}$ \emph{via} the two latter methods, \emph{\textbf{random}} or \emph{\textbf{random\_kl}} generate a sequence of CG representations of the system at a fixed degree of coarse-graining in which the $N$ sites are randomly displaced throughout the molecular structure, evaluating the associated mapping entropies. The results of this analysis can then be histogrammed along the $S_{map}$ axis, and the characteristic scale of information loss at the desired value of $N$ can be extracted, e.g., from the average mapping entropy $\mu_N$ and standard deviation $\sigma_N$ of the sample. Finally, the quality of the proposed CG representation $\mathbf{M}_N$ can be gauged in terms of its relative information gain/loss with respect to what would on average arise at the same degree of coarse-graining by randomly choosing the location of the sites; as an example, in Ref.~\citenum{giulini2020information} the chosen quantitative quality measure was the standard score $Z(\mathbf{M}_N)$ of the optimal mapping with respect to the distribution of randomly extracted ones, with
\begin{equation}
\label{eq:zscore}
Z(\mathbf{M}_N)=\frac{S_{map}(\mathbf{M}_N)-\mu_N}{\sigma_N}.
\end{equation}

We underline that all quantities appearing in Eq.~\ref{eq:zscore} should be calculated always via the same $S_{map}$ estimator in a consistent manner, that is, either through a combination of \emph{\textbf{measure}} and \emph{\textbf{random}} to respectively determine $S_{map}(\mathbf{M}_N)$ and $(\mu_N,\sigma_N)$, or alternatively through a combination of \emph{\textbf{measure\_kl}} and \emph{\textbf{random\_kl}}.

\noindent {\bf CG MAPPING OPTIMISATION —} With these instruments at hand, we now have all the necessary ingredients to introduce the last $S_{map}$-based analysis tools implemented in EXCOGITO, namely the ones devoted to the identification of the CG representations of the system of interest that, despite a reduction in resolution, are capable of preserving the largest amount of information about the statistical properties of the atomistic reference. Hence, among all the possible decimation mappings that can be designed for a macromolecule at a fixed degree of coarse-graining $N$, we are now looking for those that \emph{minimize} $S_{map}$; crucially, as it will be discussed in the following, these maximally informative CG representations appear to be related to the system's functional regions, suggesting the mapping entropy optimization workflow as a promising approach to extract relevant macroscopic insight from raw, all-atom data. 

In principle, one could consider detecting the aforementioned optimized representations by comprehensively exploring the mapping space of the system at the desired value of $N$; subsequently, one can rank the resulting possible selections of CG sites according to the information loss they generate. However, for a macromolecule composed of $n$ atoms, the number of decimation mappings retaining $N<n$ sites to be probed in this scheme would be $n!/N!(n-N)!$, which rapidly increases with the number of microscopic constituents \cite{menichetti2021journey}. This makes an exhaustive sampling approach unfeasible for all but the smallest systems; an alternative procedure is thus necessary to tackle such a high-dimensional optimization problem, where the intrinsically discrete nature of decimated CG representations also prevents, e.g., the use of gradient-based methods.

EXCOGITO enables the identification of the maximally informative CG representation of a system via the \emph{\textbf{optimize}} and \emph{\textbf{optimize\_kl}} tools, which respectively build on the equilibrium and KL mapping entropy estimators reported in Eqs.~\ref{eq:smap_beta} and ~\ref{eq:smap_kl}. Both protocols minimize $S_{map}$ by relying on a Monte Carlo simulated annealing (SA) approach, gradually pushing a stochastic exploration of the mapping space of the system to visit CG representations characterized by a low information loss, see Sec.~\ref{sec:sumtools} for all technical details. More specifically, starting from an initial selection of $N$ sites of the macromolecule, \emph{\textbf{optimize}} and \emph{\textbf{optimize\_kl}} perform a sequence of Monte Carlo moves that, at each step, propose a swap between a retained and neglected atom in the CG mapping, hence working at a fixed degree of coarse-graining. The moves are accepted according to a Metropolis-like criterion that employs $S_{map}$ as a cost function, and in which the associated ``temperature'' is exponentially decreased in the course of the simulation, driving the sampling, after an initial transient, to converge towards a local minimum of the mapping entropy. A single SA run performed with one of the two methods thus enables the identification of one of the sought-for maximally informative CG representations that can be employed to describe the system of interest.

The next necessary step in this analysis is to account for the fact that the \emph{manifold of solutions} to the optimization problem can have a rather complex structure. Indeed, given the intricacy of the network of interactions among the system's microscopic constituents, it is reasonable to expect a rugged landscape of information loss throughout the mapping space, exhibiting a whole ensemble of more or less degenerate local minima either living in relatively flat basins or being widely separated by high $S_{map}$ barriers. None of these minima can \emph{a priori} be preferred over another; rather, in order to gather a full picture of the link between resolution reduction and information content, one needs to simultaneously consider a \emph{pool of solutions} minimising the mapping entropy. As we will discuss in the following, it is precisely the pattern that emerges from the analysis of the whole ensemble of such optimized solutions that enables the extraction of nontrivial information about the system and its function.

With reference to the software, the \emph{\textbf{optimize}} and \emph{\textbf{optimize\_kl}} tools of EXCOGITO enable the user to tackle the simultaneous extraction of several (local) minima of the mapping entropy through multiple SA runs that are performed in parallel, see Sec. \ref{sec:sumtools} for all technical details.

\subsection{Metric in the coarse-grained mapping space and related EXCOGITO tools}
\label{sec:mapspacetools}

Equations \ref{eq:smap_beta} and \ref{eq:smap_kl} allow us to calculate the mapping entropy associated to a given mapping, and their optimisation leads to the identification of mappings that entail the largest amount of information about the system for a given number of retained atoms. It is also instructive, however, to broaden the perspective on mappings themselves, and to investigate their properties from a purely structural perspective.

More specifically, given a mapping as the selection of a particular subgroup of elements (the retained atoms) from a set (the whole molecule), we can ask ourselves questions about the total number of mappings, the amount of them sharing the same qualitative and quantitative features, and the relationship that exists between mapping groups with given features and the structural properties of the system on which the selection takes place. These questions have been addressed in various recent works \cite{foley2020exploring,menichetti2021journey,kidder2024surveying}; here, we only report those results obtained by some of us \cite{menichetti2021journey} through the application of the methods implemented in EXCOGITO.

In order to assess even the simplest properties of the mapping space associated to a given protein, one needs to define basic quantitative instruments, for example to determine how different a given mapping is from another. To this end, some of us introduced a notion of norm, cosine, and distance between CG mappings, which only make use of the structural features of the biomolecule of interest. More specifically, we define the scalar product $\langle {\bf M},{{\bf M}'}\rangle$ between two mappings ${\bf M}$ and ${\bf M}'$ as:
\begin{equation}
\label{eq:scal_prod}
\langle {\bf M},{{\bf M}'}\rangle=\sum_{i,j=1}^n e^{-r^2_{ij}/4\sigma^2}\chi_{{\bf M},i}\chi_{{\bf M}',j},
\end{equation}
where $\chi_{{\bf M},j}$ is the mapping function as defined in Eq.\ref{eq:mapping}, $r_{ij}$ is the distance between atoms $i$ and $j$, and $\sigma$ is a parameter that tunes the amplitude of the Gaussian employed to calculate the \emph{coupling} $J_{ij} = e^{-r^2_{ij}/4\sigma^2}$ between them. In the work in which it was introduced, this parameter was set to $\sigma = 0.19$ nm, however its value can be tuned according to the specific application of interest.

From Eq. \ref{eq:scal_prod} we can calculate the (squared) norm of a mapping as
\begin{equation}
\label{eq:norm_gauss}
\mathcal{E}({\bf M})=\langle {\bf M},{\bf M}\rangle=\sum_{i,j=1}^n J_{ij}\ \chi_{{\bf M},i}\ \chi_{{\bf M},j}.
\end{equation}

Having defined a scalar product and a norm, we can then introduce the cosine between two mappings ${\bf M}$ and ${\bf M}'$,
\begin{equation}
\label{eq:cosine}
\cos\theta_{{\bf M},{\bf M}'}=\frac{\langle {\bf M},{{\bf M}'}\rangle}{\left(\mathcal{E}({\bf M})\mathcal{E}({\bf M}')\right)^{\frac{1}{2}}},
\end{equation}
as well as the distance between two mappings:
\begin{eqnarray}
\label{eq:dist_dotmapp}
&&\mathcal{D}({\bf M},{\bf M}')= \left(\mathcal{E}({\bf M})+\mathcal{E}({\bf M}')-2\langle {\bf M},{{\bf M}'}\rangle\right)^{\frac{1}{2}}\nonumber \\
&&=\left( \sum_{i,j=1}^n J_{ij}\ \chi_{{\bf M},i}\ \chi_{{\bf M},j} \;+\; \sum_{i,j=1}^n J_{ij}\ \chi_{{\bf M}',i}\ \chi_{{\bf M}',j}\; -2\sum_{i,j=1}^n J_{ij}\ \chi_{M,i}\ \chi_{{\bf M}',j}\right)^{\frac{1}{2}}.
\end{eqnarray}

In Ref.~\citenum{menichetti2021journey} the norm and the distance were rescaled by a function of the atomistic coordination number:
\begin{equation}
\label{eq:at_coord}
\bar{z}=\frac{1}{n}\sum_{i,j=1}^n J_{ij},
\end{equation}
calculated over a specific molecular configuration (e.g. the initial frame of an MD trajectory, or the frame closest to the average conformation of the system). Consequently, $\mathcal{E}({\bf M})$ and $\mathcal{D}({\bf M},{\bf M}')$ read
\begin{equation}
\label{eq:rescaled_norm}
\mathcal{E}_{\bar{z}}({\bf M})~=\frac{1}{\bar{z}}~\mathcal{E}({\bf M}),
\end{equation}
\begin{equation}
\label{eq:rescaled_dist}
\mathcal{D}_{\bar{z}}({\bf M},{\bf M}')=\frac{1}{\sqrt{\bar{z}}}~\mathcal{D}({\bf M},{\bf M}'),
\end{equation}
while the formula for the cosine remains unaltered. This normalisation accounts for the specificity of the structural features of the protein, sets a characteristic scale for the value of scalar product, and enables a fair comparison between mapping pairs defined on molecules of different size \cite{menichetti2021journey}.

\begin{figure*}
\centering
                \includegraphics[width=.9\textwidth]{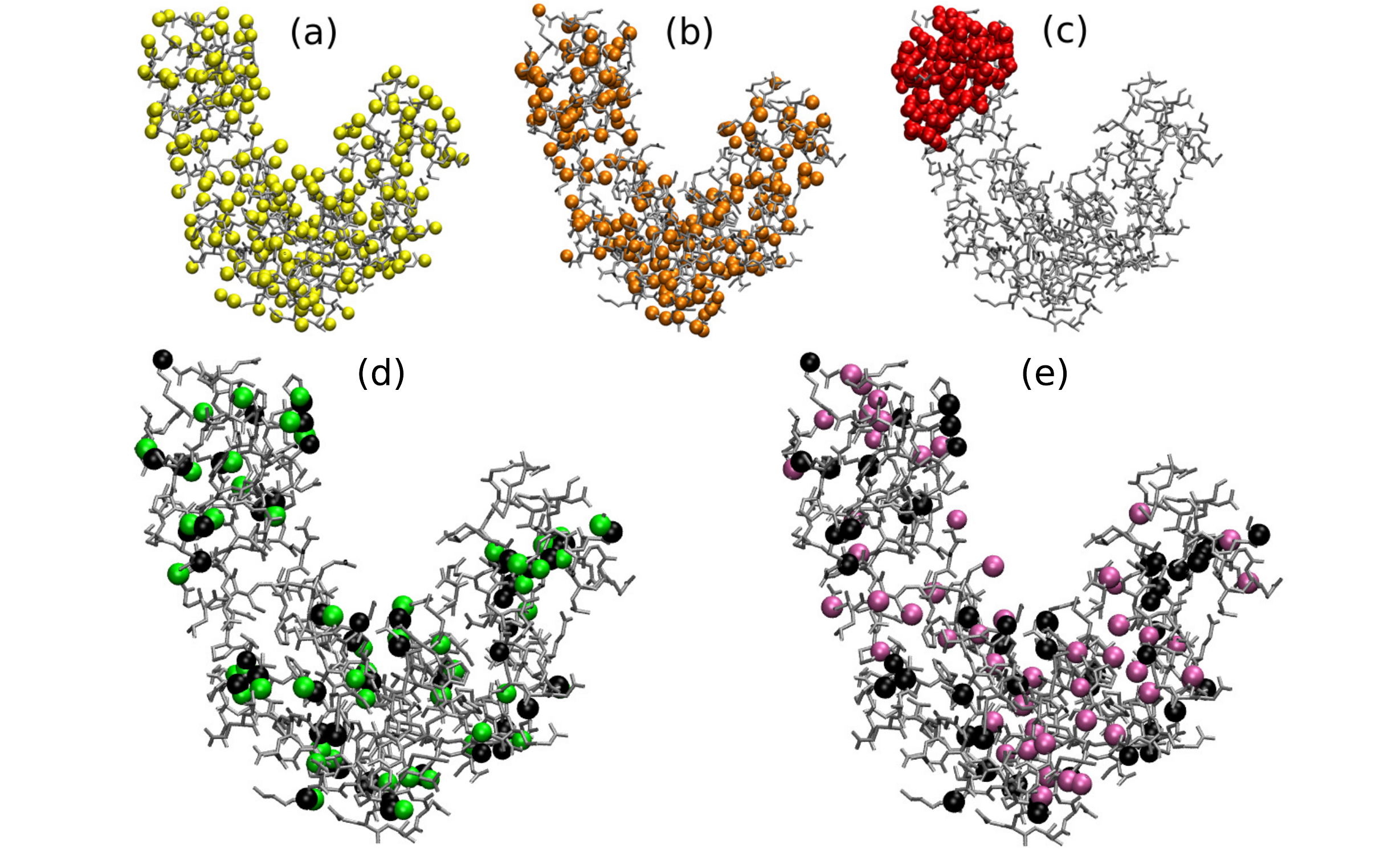}
                \caption{\label{fig:norm_and_cosines} Panels (a) to (c): examples of low, intermediate, and high norm mappings, respectively, for adenylate kinase. Panels (d) and (e): parallel and orthogonal mappings, respectively. Figure reproduced from: Menichetti  \emph{et al.} \cite{menichetti2021journey}, ``A journey through mapping space: characterising the statistical and metric properties of reduced representations of macromolecules'', Eur. Phys. J. B 94, 204 (2021). This image is licensed under CC BY 4.0 \href{http://creativecommons.org/licenses/by/4.0}{http://creativecommons.org/licenses/by/4.0}.}
\end{figure*}

The norm of a mapping, or equivalently its square in Eq. \ref{eq:rescaled_norm}, is a measure of the mapping's ``compactness'': it was observed, in fact, that CG representations with high value of the norm select groups of atoms that are very close to each other, hence corresponding to coarse pictures in which a relatively compact region of the molecule is represented with high resolution, while the remainder is largely discarded. In contrast, low-norm mappings correspond to very sparse selections, in which the retained atoms are maximally distant one from the other compatibly with their number and the properties of the molecule. Illustrative examples of these cases are reported in Fig. \ref{fig:norm_and_cosines} (panels $a$ to $c$).

Furthermore, mappings can be more or less \emph{parallel}: a high scalar product between mappings indicates that the atom selections under examination are largely similar; this can either mean that both of them retain at least in part the same atoms, or that the selected atoms of a mapping are very close to those of the other.
In Fig. \ref{fig:norm_and_cosines} (panels $d$ and $e$) we show a pair of parallel mappings and a pair of orthogonal ones: it is interesting to observe that the difference between the two cases is hard to grasp by eye, however the distance between the atoms in the two pairs of selections is on average very low in the first case, and rather high in the second.

In the following sections, we illustrate the application of the mapping metrics presented insofar to a particular case study. Further details on the mapping norm, cosine etc. can be found in Ref.~\citenum{menichetti2021journey}.

\subsection{Summary of the EXCOGITO tools}\label{sec:sumtools}

\begin{figure}
\centering
    \includegraphics[width=\columnwidth]{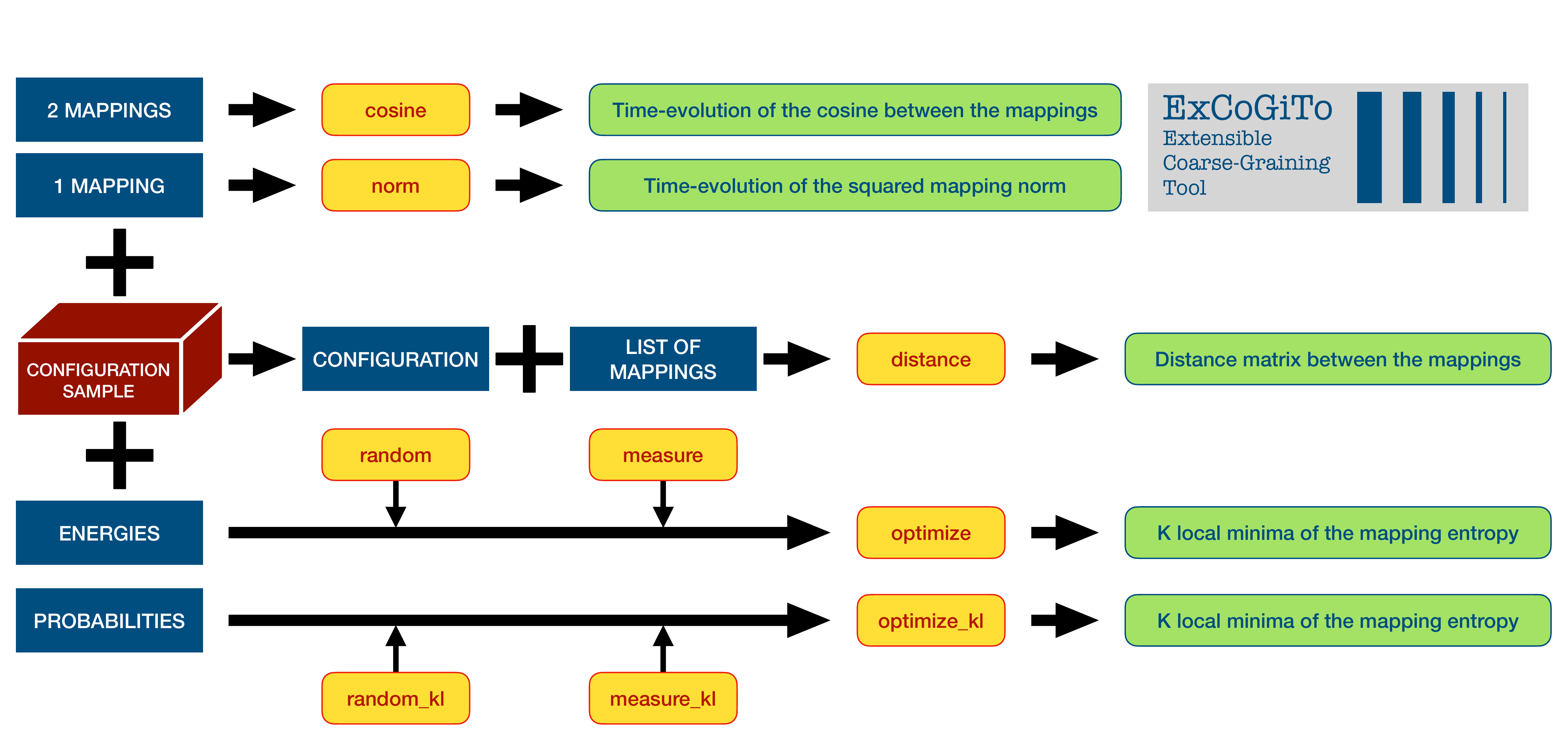}
\caption{\label{fig:scheme} Scheme of the EXCOGITO methods, inputs, and outputs. The starting point is a sample of molecular configurations, or a given structure thereof; these can be obtained from an equilibrium MD trajectory, but other sources (e.g. non-equilibrium MD simulations, ensembles of experimental structures) are possible. Adding the potential energy (in the case of equilibrium sampling) or the probability associated with each frame one can estimate the value of the mapping entropy of a given CG representation, generate a number of random mappings, and eventually optimise $S_{map}$ through simulated annealing. Alternatively, the trajectory can be analysed through the computation of the norm of a given CG mapping and/or the scalar product between two or more mappings throughout the frames.}
\end{figure}

At present, EXCOGITO contains the subprograms listed and described hereafter. The organisation of these methods and their relationship, input, and output is graphically illustrated in Fig. \ref{fig:scheme}. Technical aspects of the execution of EXCOGITO are provided in Sect. \ref{sec:launchexcogito} of the Appendix, where we list the command line syntax of the various methods in Tab. \ref{tab:syntax} as well as the parameters required in input in Tab. \ref{tab:params}.

\begin{itemize}[leftmargin=0pt]

\item \emph{\textbf{measure}}: the user provides in input to EXCOGITO a series of configurations extracted from an equilibrium MD simulation of the system of interest, the associated energies, and a specific choice of the CG mapping in the form of a text file---a prototype is available in the examples. The associated mapping entropy $\tilde{S}_{map}$ is then computed according to Eq.~\ref{eq:smap_beta};

\item \emph{\textbf{measure\_kl}}: Kullback-Leibler-based routine for the computation of the mapping entropy \emph{via} Eq. \ref{eq:smap_kl}.  Here, the user provides EXCOGITO with a set of atomistic configurations, the associated probabilities, as well as a CG mapping;

\item \emph{\textbf{random}}: generation of \texttt{n\_mappings} random CG mapping of the system with a fixed number of \texttt{cgnum} sites (see Tab.~\ref{tab:params}) and measurement of the corresponding values of $\tilde{S}_{map}$. This task is useful, e.g., when one wants to compare the values of $\tilde{S}_{map}$ of optimal mappings to those of CG representations that are randomly drawn from the mapping space;

\item \emph{\textbf{random\_kl}}: Kullback-Leibler version of task \emph{\textbf{random}}, making use of the $\hat{S}_{map}$ estimator;

\item \emph{\textbf{optimize}}: optimization task that produces $K$ local minima of the mapping entropy $\tilde{S}_{map}$ in the space of the possible CG representations of the system that retain \texttt{cgnum} sites (see Tab.~\ref{tab:params}). The number of minima $K$ has to be lower or equal to the number of CPU cores of the employed architecture since each core performs a single minimization. The algorithm employed for the optimization is Monte Carlo simulated annealing; at each step, the current mapping ${\bf M}$ is slightly modified into a new one ${\bf M}'$ by replacing a retained atom with another one that was not part of ${\bf M}$. Such a move is accepted or rejected with a probability given by a Metropolis criterion:
\begin{equation}
\label{eq:wl_accept}
    W({\bf M} \rightarrow {\bf M}') = \text{min} \left[1,e^{(\tilde{S}_{map}({\bf M}) - \tilde{S}_{map}({\bf M}'))/T}\right], 
\end{equation}
in which the simulated annealing temperature $T$ experiences an exponential decay in time dictated by
\begin{equation}\label{eq:sat}
T(i) = T_0\ e^{-i/\nu},
\end{equation}
where $i$ is the considered optimization step and $\nu$ tunes the amplitude of the decay. The user can choose the overall number of MC steps, together with $T_0$ and $\nu$ of Eq. \ref{eq:sat} (see Tab.~\ref{tab:params}).

\item \emph{\textbf{optimize\_kl}}: analogous to \emph{\textbf{optimize}}, but using the Kullback-Leibler version of the mapping entropy $\hat{S}_{map}$;

\item \emph{\textbf{norm}}: given a CG representation and a trajectory, the time-evolution of the squared mapping norm in Eq.~\ref{eq:rescaled_norm} is calculated. The value of the atomistic coordination number (Eq.~\ref{eq:at_coord}) is chosen as the one calculated over the first structure provided in input;

\item \emph{\textbf{cosine}}: given two CG representations and a trajectory, the time-evolution of the cosine (Eq.~\ref{eq:cosine}) between them is calculated;

\item \emph{\textbf{distance}}: given a set of \texttt{n\_mappings} (see Tab.~\ref{tab:params}) CG representations and a given configuration of the molecule, the distance matrix between them is computed using Eq.~\ref{eq:rescaled_dist}. Such matrix can be employed for several purposes, such as the calculation of the sketch maps as in Ref.~\citenum{menichetti2021journey}.
\end{itemize}

\section{Previous applications of EXCOGITO and software performance}

In this section, we report a selection of the applications of EXCOGITO from previous works. We begin by addressing the calculation and minimisation of the mapping entropy; subsequently, we discuss the usage of the mapping space metric tools; lastly, we provide information on the performance of the software in a few selected cases, as indicative guidelines on the running time needed for performing the tasks of EXCOGITO.

\subsection{Applications of the mapping entropy tools}
\label{sec:prevappsmap}

\begin{figure*}
\centering
                \includegraphics[width=.9\textwidth]{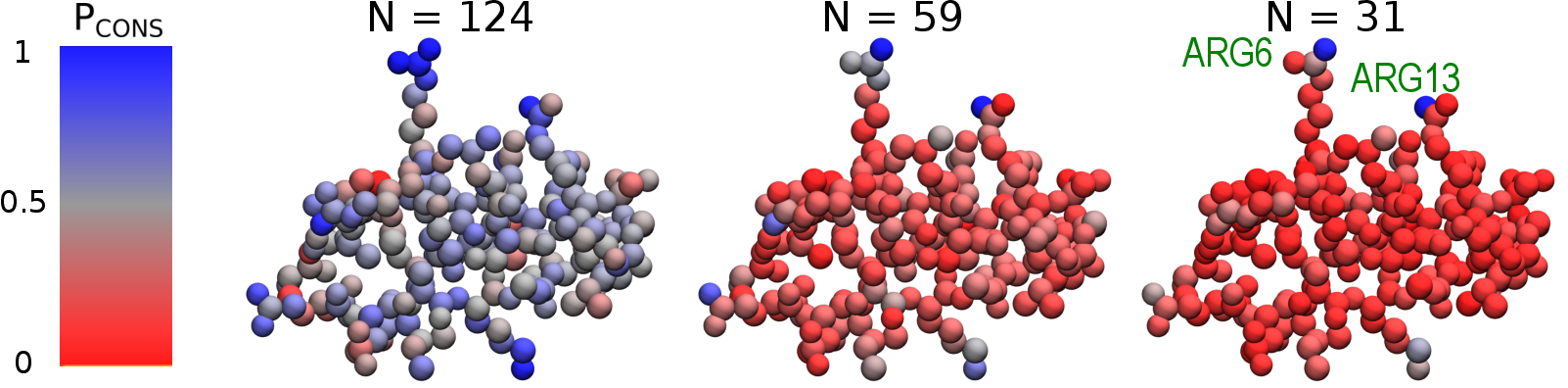}
                \caption{\label{fig:tamapin_atoms_map} Structure of Tamapin colored according to the probability of preserving an atom in the pool of EXCOGITO optimised mappings. Figure reproduced from: Giulini \emph{et al.} \cite{giulini2020information}, ``An Information-Theory-Based Approach for Optimal Model Reduction of Biomolecules'', J. Chem. Theory Comput. 2020, 16, 11, 6795–6813, \href{https://pubs.acs.org/doi/full/10.1021/acs.jctc.0c00676}{https://pubs.acs.org/doi/full/10.1021/acs.jctc.0c00676}. Further permissions related to this figure should be directed to the ACS.}
\end{figure*}

The work by Giulini and coworkers ~\cite{giulini2020information} describes the first application of the mapping entropy optimisation workflow to three markedly different proteins, namely the tamapin toxin, adenylate kinase, and $\alpha$-$1$ antitrypsin. Upon simulating these molecules in explicit solvent, the distribution of values of $\tilde{S}_{map}$ is calculated on $500$ randomly selected mappings (\emph{\textbf{random}} subcommand). Then, several optimisations are run (\emph{\textbf{optimize}} subcommand), resulting in mappings that correspond to local minima of $\tilde{S}_{map}$. Upon averaging over these solutions, it is evident how the mapping entropy optimisation assigns an uneven level of detail to the structures, with some amino acids that are retained more often than others. In all the three considered cases, the retained amino acids are heavily involved in the biological role of the protein, and in particular in the binding to another molecule \cite{giulini2020information}. This is a consequence of the fact that, in simulations performed in absence of the substrate, amino acids involved in the binding tend to correlate with important energetic fluctuations at the level of the whole protein, and feature high conformational variability. These two characteristics determine a higher chance for these residues to be retained in an optimal representation: in fact, the knowledge of their position and arrangement provides a better picture of the system as a whole at a lower resolution level, in comparison to other residues whose structural properties are less informative.

Fig.~\ref{fig:tamapin_atoms_map} shows an example of this behavior for the tamapin toxin: the minimisation of the mapping entropy for various degrees of resolution (i.e. for different numbers of retained atoms) consistently leads to the conservation of terminal atoms of ARG6 and ARG13, which are the two amino acids responsible for the binding to the toxin's substrate, the SK2 calcium-activated potassium channel \cite{giulini2020information}.

\subsection{Applications of the mapping space metric tools}
\label{sec:prevappmetric}

The notions of norm, cosine, and distance for coarse-grained representations have been introduced in Ref.~\citenum{menichetti2021journey}, where these basic quantities have been employed to characterise the metric properties of the mapping space of specific proteins. In that work, it was shown that the mapping space is extremely diverse; that the mappings in it can be grouped according to features that correlate with the structure of the underlying protein; and that in this space a phase transition occurs, that is analogous to a gas-liquid phase transition on the lattice, as it was observed by other authors as well \cite{foley2020exploring, kidder2024surveying}.

Recently, the concepts of mapping space metric have been applied by Giulini and coworkers to the analysis of interface residues in protein complexes \cite{giulini2024arctic}. By exploiting the equivalence between protein-specific interfaces and coarse-grained mappings, Eqs.~\ref{eq:cosine} and ~\ref{eq:dist_dotmapp} can be used to quantify the similarity between different interfaces and to cluster them in binding surfaces.

\subsection{Performance and optimisation}
\label{sec:performance}

The runtime of EXCOGITO largely depends on several factors. First, the various tasks have very different computational requirements, as some of them consist of rapid one-time calculations (e.g., individual computations of \emph{\textbf{norm}}, \emph{\textbf{cosine}}, or \emph{\textbf{measure}}, such tasks being listed in order of increasing effort), while others (most notably \emph{\textbf{optimize}} and \emph{\textbf{optimize\_kl}}) are iterative, and thus intrinsically much more time-consuming. Furthermore, also the system of interest affects the performance, most notably through its size (number of atoms/residues of the protein) and the number of configurations passed as input to the software. Finally, additional ingredients that influence the performance of EXCOGITO are, quite naturally, the architecture and efficiency of the machine it is run on, as well as the compilation options. The latter, in particular, can considerably speed up the calculations; to this end, a few general guidelines to achieve this boost through compiler optimisation are provided in the README of the software, thereby leaving the users the opportunity to improve the efficiency in the most appropriate manner for their equipment.

From the preceding discussion, it follows that a comprehensive assessment of the computational effort required to perform all the possible EXCOGITO tasks, further covering different combinations of the input parameters, would be difficult to achieve. To help the users roughly estimate the computational and time resources needed for performing their analyses, we here focus our attention on the computing time associated with the fundamental building block of EXCOGITO, namely the \emph{\textbf{measure}} tool calculating the $\tilde{S}_{map}$ associated the choice of a specific CG representation of the system, see Eq.~\ref{eq:smap_beta}. Critically, this task lies at an intermediate level of computational effort among all the EXCOGITO ones.

The average runtime necessary for performing a single estimate of $\tilde{S}_{map}$ for a random CG mapping via \emph{\textbf{measure}} is presented in Table~\ref{tab:perf} for three biomolecules of increasing size, namely tamapin, adenylate kinase, and $\alpha$-$1$-antitrypsin.\cite{giulini2020information} In each of the three cases, we fix the number of CG sites to be equal to the number of residues in the protein and resort to the standard clustering criterion implemented in EXCOGITO for extracting the CG macrostates (\texttt{criterion = 0} in Sec.~\ref{sec:confcl}), reporting results for different numbers of configurations provided in input to the software. For the largest number of frames ($10^4$), the computational times required to perform the calculations via the standard clustering method on the three systems are further compared to those obtained by relying on the fast clustering strategy (\texttt{criterion = 3}, \texttt{stride = 3}); as discussed in Sec.~\ref{sec:confcl}, the latter can be employed only provided that a time-continuous MD trajectory of the system is available. All computations were performed on a single core of an Intel Xeon-Gold 5118 processor, compiling the software in a \emph{plain vanilla} setup without any optimisation. Overall, in the case of the standard clustering, we observe that for a fixed number of input configurations the runtime of EXCOGITO scales linearly with the number of (heavy) atoms in the system. On the other hand, for a fixed number of constituent atoms, the computational time scales quadratically with the number of frames. Finally, we note that employing the fast clustering method can speed up $S_{map}$ roughly by a factor of 10. As anticipated, we stress that the variables that can affect the program performance are many, hence the real runtime can vary with respect to the estimates reported in Table~\ref{tab:perf}.

\begin{table}[h]
    \centering
    \begin{tabular}{lccccccc}
    \toprule
         \multirow{2}{*}{Protein} & \multirow{2}{*}{$n_{at}$} & \multirow{2}{*}{$N_{CG}$} & \multicolumn{5}{c}{CPU time $(s)$}\\
        \cmidrule(lr){4-8}         
        & & & $f=1000$ & $f=2500$ & $f=5000$ & $f=10000$ & F.C.  \\ \midrule
         \addlinespace[0.1cm] TAM & 230 & 31 & 1 & 5.8(4) & 25(2) & 105(7) &12(1)\\
        \hline
         \addlinespace[0.1cm]
         AKE & 1656 & 214 & 7.1(7) & 41(3) & 160(4) & 631(8) & 74(1)\\ 
         \hline
          \addlinespace[0.1cm]
         AAT & 2956 & 372 & 12.7(6) & 72(2) & 279(2) & 1102(4) & 130(2)\\
         \hline
    \end{tabular}
    \caption{Average CPU time (in seconds) needed to run a single $S_{map}$ estimate via the EXCOGITO \emph{\textbf{measure}} tool for three proteins with an increasing number of constituent atoms $n_{at}$---namely tamapin (TAM), adenylate kinase (AKE), and $\alpha$-$1$-antitrypsin (AAT).\cite{giulini2020information} The number of retained sites $N_{CG}$ was in each case set equal to the number of residues in the molecule, and the computational time reported was calculated as an average over $20$ randomly extracted CG mappings of the system. We display the results associated with the standard clustering method of EXCOGITO (see Sec.~\ref{sec:confcl}) for different numbers of input frames $f$ employed in the calculations, as well as, for $f=10^4$, the runtime one achieves by relying on the fast clustering strategy (column ``F.C.'', see Sec.~\ref{sec:confcl}). Computations were performed on a single core of an Intel Xeon-Gold 5118 processor.}
    \label{tab:perf}
\end{table}

\section{Example application of EXCOGITO: icosalanine}
\label{sec:appicosal}

In this section we showcase the usage of EXCOGITO through its application to a toy system, namely \textit{icosalanine}, a chain of $20$ alanine residues ($101$ heavy atoms). The system is properly equilibrated and then simulated for $200$ nanoseconds using GROMACS 2018 \cite{van2005gromacs} with the standard amber99sb-ildn forcefield \cite{lindorff2010improved}, as in Ref.~\citenum{giulini2020information}. We extract a configuration with the associated energy (calculated as described in Ref.~\citenum{giulini2020information}) every $20$ ps.

\begin{figure*}
\includegraphics[width=\columnwidth]{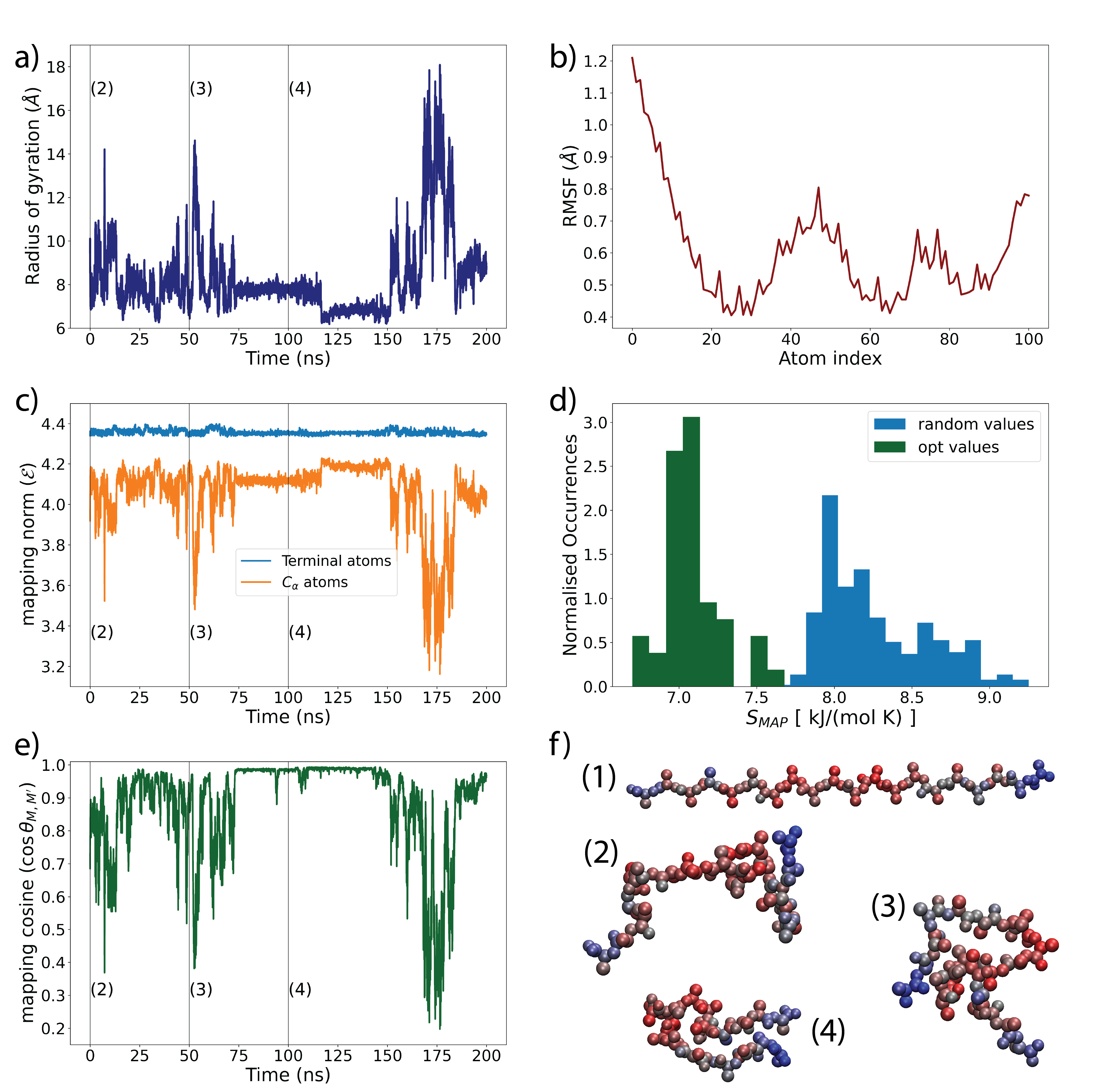}
\caption{Results of the EXCOGITO analyses performed on icosalanine. Panels (\textit{a-b}): values of the radius of gyration and root mean squared fluctuations (RMSF) extracted from an MD simulation of the system. Panel \textit{c}: time evolution of the mapping norms for the $C_{\alpha}$ and ${\bf M}^{nter}$ mappings (see main text). Panel \textit{d}: comparison between the distribution of mapping entropy values of random (blue histogram) and optimized mappings (green histogram). Panel \textit{e}: time evolution of the cosine between ${\bf M}^{nter}$ and ${\bf M}^{cter}$ (see main text). Panel \textit{f}: probability of conserving each (heavy) atom in the icosolanine structure, displayed in the original, unrealistic, fully stretched conformation (1), and in three realistic conformations observed during the simulation---partially stretched (2), partially folded (3), and mainly folded (4).}
\label{fig:polyala}
\end{figure*}

Fig.~\ref{fig:polyala}\textit{a-b} report, respectively, the radius of gyration of the system along the trajectory and the root mean square fluctuation (RMSF) of its atoms. During the simulation, the molecule is observed to undergo several conformational transitions, going back and forth between what are rather elongated, partially folded, and completely folded arrangements. This variability is made evident by the time evolution of the protein's radius of gyration displayed in Fig.~\ref{fig:polyala}\textit{a}, whose value ranges from $6$ to $18$ \AA.

Also the RMSF presented in Fig.~\ref{fig:polyala}\textit{b} is indicative of the molecule's flexibility, with its central regions and termini fluctuating importantly. Icosalanine displays a quite unstructured, polymer-like behavior, in that it broadly explores the conformational space without collapsing into a stable native structure. Evidence of this is the fact that, despite the sequence homogeneity of the molecule, its RMSF results appear to be asymmetric around the central monomers of the sequence: this is in part because the N and C termini are chemically different, and they hence entertain distinct interactions with the remainder of the molecule and the solvent. On the other hand, the $200$ ns duration of the trajectory, which generally allows a good sampling of the native state for a similarly-sized, globular protein, is probably too short for icosalanine to explore the conformational space exhaustively. However, this is not an issue in the present case, since the objective of this simulation is not to investigate the properties of icosalanine, but rather to illustrate the methods of EXCOGITO.

To demonstrate the usage of EXCOGITO in practice, we start by elucidating the behaviour of the squared mapping norm (\emph{\textbf{norm}} tool, Eq.~\ref{eq:rescaled_norm}) for two markedly different mapping operators throughout this system's trajectory, see Fig.~\ref{fig:polyala}\textit{c}. The first operator is the ${\bf M}^{C_{\alpha}}$ mapping, obtained by retaining only the $C_{\alpha}$ atoms of the system, while the second, ${\bf M}^{Nter}$, contains only the first $20$ atoms of the chain starting from the N terminal. Intuitively, the first mapping is very uniformly distributed over the protein, while the second is extremely localised in a specific region. From the plot we observe how $\mathcal{E}({\bf M}^{Nter})$ is consistently higher than $\mathcal{E}({\bf M}^{C_{\alpha}})$, as expected given its higher globularity. Moreover, $\mathcal{E}({\bf M}^{Nter})$ does not display relevant fluctuations, as the atoms retained by ${\bf M}^{Nter}$ do not change their mutual distances appreciably during the simulation. Instead, the ${\bf M}^{C_{\alpha}}$ mapping induces very wide fluctuations in $\mathcal{E}$, due to the continuous folding and unfolding of the polypeptide.

As for the geometrical relation among different CG representations, Fig.~\ref{fig:polyala}\textit{e} shows the cosine (\emph{\textbf{cosine}} tool, Eq.~\ref{eq:cosine}) of the angle between two mappings with $N=20$, namely ${\bf M}^{Nter}$ (see above) and ${\bf M}^{Cter}$, which contains only atoms coming from the C terminal region of the peptide. The cosine is very low when the polypeptide is in a stretched conformation and the mappings are therefore almost orthogonal. Instead, it approaches $1.0$ when the polypeptide is in a packed conformation (such as Fig.~\ref{fig:polyala}\textit{f} (4)) and atoms of the two mappings are very close to each other, giving rise to an almost perfect parallelism.

Finally, we quantify (\emph{\textbf{measure}} tool) the mapping entropy for $500$ randomly extracted mappings with $40$ CG sites. We then follow the standard simulated annealing protocol (\emph{\textbf{optimize}} tool) to minimize $S_{map}$ over $48$ independent optimizations. Fig.~\ref{fig:polyala}\textit{d} reports the non-overlapping distributions of values of $S_{map}$ arising for the two sets of mappings, and Fig.~\ref{fig:polyala}\textit{f} shows how the probability of retaining each atom in the optimized solutions is unevenly distributed over the polypeptide chain. The two terminal regions are highly conserved by the minimization procedure, while the central region is more coarse-grained, especially in its $C_{\beta}$ atoms. Given the analysed set of configurations and values of energy, this suggests that the optimal CG mapping should assign higher resolution to the two terminal regions of the peptide, with a coarser description of the central region that only retains some backbone atoms.

\section{Conclusions}
\label{sec:concl}

Recent works \cite{diggins2018optimal, giulini2020information, foley2020exploring, giulini2021system, kidder2024surveying} have emphasized the fundamental importance of the mapping between high- and low-resolution descriptions of a system, whose origins are to be found in the renormalisation group approach to critical phenomena. While significant efforts have been directed towards developing CG force fields, one can observe in the literature a notable gap in the examination of the properties of mappings themselves, which nonetheless a few authors have started to address. The concept of mapping entropy, defined as the divergence between the reference, high-resolution configuration distribution and the one reconstructed from a low-resolution representation, has emerged as an important measure of the information retained by a mapping. The minimization of mapping entropy, in fact, offers valuable insights into the system's function, unveiling nontrivial knowledge about its physical, chemical, and potentially biological properties. Furthermore, mappings can be leveraged even in the absence of sampled conformations, making use of \emph{ad hoc} metrics to identify structural features and quantitatively explore the mapping space.

In this work we have presented EXCOGITO, a suite of routines that enables the analysis of macromolecular systems making use of the properties of mappings, most notably by means of quantities such as mapping entropy and mapping metrics. EXCOGITO is a toolbox software platform implemented in C and freely available from a public repository, and stands as a user-friendly tool empowering researchers to effectively explore the properties of complex biological or artificial macromolecules through the lens of low-resolution representations. By making this software available to the community, we hope to contribute to the field of soft and biological matter modelling, and facilitate further advancements in understanding complex molecular systems.

\section{Appendix}

\subsection{Launching EXCOGITO: mandatory files and external parameters}
\label{sec:launchexcogito}

The {\sc README} file of EXCOGITO provides all the necessary details to compile and run the calculations. In addition, the PDF documentation created with \textit{doxygen} is available in the \textit{docs}.

Each task of EXCOGITO can be launched from the command line using the syntax reported in Tab.~\ref{tab:syntax}. A mandatory argument for each subprogram is the \textit{ini} parameter file (pfile in Tab.~\ref{tab:syntax}), which contains the necessary hyperparameters that must be provided by the user in order to run the desired task. A list of the available parameters, together with a short explanation of their role, is available in Tab.~\ref{tab:params}.

\begin{table*}
\centering
\caption{\label{tab:syntax}How to launch EXCOGITO subprograms?}
\begin{tabular*}{\textwidth}{ll}
\toprule
\textbf{Subprogram} & \textbf{Syntax} \\
\midrule
optimize & \small{excogito optimize -p pfile.ini -t tfile.xyz -e energies.txt -c code}\\
optimize\_kl & \small{excogito optimize\_kl -p pfile.ini -t tfile.xyz -e probs.txt -c code}\\
random & \small{excogito random -p pfile.ini -t tfile.xyz -e energies.txt -c code}\\
random\_kl & \small{excogito random\_kl -p pfile.ini -t tfile.xyz -e probs.txt -c code}\\
measure & \small{excogito measure -p pfile.ini -t tfile.xyz -e energies.txt -c code -m mapping.txt}\\
measure\_kl & \small{excogito measure\_kl -p pfile.ini -t tfile.xyz -e probs.txt -c code -m mapping.txt}\\
norm & \small{excogito norm -p pfile.ini -t tfile.xyz -c code -m mapping.txt}\\
cosine & \small{excogito cosine -p pfile.ini -t tfile.xyz -c code -m mapping.txt -n mapping2.txt}\\
distance & \small{excogito distance -p pfile.ini -t tfile.xyz -c code -x mapping\_matrix.txt}\\
\bottomrule
\end{tabular*}
\begin{tablenotes}%
\item Each EXCOGITO subprogram requires a set of input files and codes, each one denoted with a letter. As an example, the parameter file must be preceded by a \quotes{-p}. The input elements that are always mandatory for EXCOGITO are the \textit{ini} parameter file (pfile), the \textit{xyz} trajectory file (tfile) and the \textit{code} string, employed to create output files. The flag \quotes{-e} accepts a file containing the energy (here generally indicated with energies.txt) or the probability of each microstate (probs.txt).
\end{tablenotes}
\end{table*}

Another mandatory argument for each subprogram is an \textit{xyz} trajectory file containing \texttt{frames} (see Tab.~\ref{tab:params}) sampled configurations of the biomolecular system of interest, viewed at the atomistic level. The \textit{xyz} format of the trajectory file must follow this syntax:
\begin{lstlisting}[language=bash] 
230
string1
string2   25.380   20.910   35.540
string2   25.790   19.570   35.120
\end{lstlisting}
The first number must be equal to \texttt{atomnum} (see Tab.~\ref{tab:params}), the number of atoms in the atomistic trajectory. string1 and string2 are custom strings that can be used to annotate the name of the biomolecule (string1) and the atomic chemical properties (string2).

\begin{table*}
\tabcolsep=0pt
\begin{threeparttable}
\begin{tabular*}{\textwidth}{@{\extracolsep{\fill}}llccc@{\extracolsep{\fill}}}
\toprule%
Parameter name & Description & Type & \small{Mandatory} & Suggested value \\
\midrule
\texttt{frames} & number of frames in the trajectory & int & all & $< 10000$\tnote{1}\\
\texttt{atomnum} & number of atoms in the system & int & all & \\
\texttt{cgnum} & number of CG sites & int & all & $<\texttt{atomnum}$ \\
\texttt{nclust} & number of CG macrostates & int & C0 - C3 & $\in [\frac{\text{\texttt{frames}}}{500} , \frac{\text{\texttt{frames}}}{100}] $\\
\texttt{n\_mappings} & number of mappings & int & R-D & \\
\texttt{MC\_steps} & number of SA steps & int & O & $\in [5000,20000]$\\
\texttt{rotmats\_period} & SA steps between two alignments & int & O &\\
\texttt{t\_zero} & $T_0$ (Eq.~\ref{eq:sat}) for optimization tasks& float & no &\\
\texttt{criterion} & clustering criterion & int & O-R-M & \\
\texttt{distance} & cophenetic distance threshold & float & C1 &\\
\texttt{max\_nclust} & higher number of clusters & int & C2 & $\in [\frac{\text{\texttt{frames}}}{100} , \frac{\text{\texttt{frames}}}{50}] $\\
\texttt{min\_nclust} & lower number of clusters & int & C2 & $\in [\frac{\text{\texttt{frames}}}{500} , \frac{\text{\texttt{frames}}}{1000}] $\\
\texttt{Ncores} & number of cores to employ & int & no &\\
\texttt{decay\_time} & temperature decay in SA ($\nu$, Eq.~\ref{eq:sat}) & float & no &\\
\texttt{rsd} & use rsd instead of rmsd & int & no & \\
\texttt{stride} & distance between pivot conformations & int & C3 & $[2,10]$\\
\bottomrule
\end{tabular*}
\begin{tablenotes}%
\item [1]if \texttt{criterion} $\neq 3$. In that case one must consider \text{\texttt{frames}}/\text{\texttt{stride}}.
\item List of parameters of EXCOGITO. In the mandatory column, \textit{all} (resp. \textit{no}) indicates parameters that are always (resp. never) mandatory, while O, R, M, and D refer to parameters that are mandatory only for \emph{\textbf{optimize}}, \emph{\textbf{random}}, \emph{\textbf{measure}}, and \emph{\textbf{distance}} (including the \emph{\textbf{\_kl}} counterparts) tasks, respectively. C0, C1, C2, C3, C4 correspond to the different clustering criteria (Sec.~\ref{sec:confcl}): for example, if the selected \texttt{criterion} is $2$, parameters \texttt{min\_nclust} and \texttt{max\_nclust} must be present.
\end{tablenotes}
\end{threeparttable}
\caption{\label{tab:params}Available input parameters of EXCOGITO}
\end{table*}

\subsection{{\bf Comparing coarse-grained structures and clustering the conformational space}}\label{sec:confcl}

The subprograms \emph{\textbf{optimize}}, \emph{\textbf{optimize\_kl}}, \emph{\textbf{random}}, \emph{\textbf{random\_kl}}, \emph{\textbf{measure}}, and \emph{\textbf{measure\_kl}} require the definition of a set of CG macrostates ${\bf R}$ out of the original microstates ${\bf r}$ of the atomistic system. The identification of these macrostates is here carried out by means of a clustering procedure that lumps together the \texttt{frames} mapped projections  of the atomistic system, i.e. the configurations of the system in terms of the subset of retained atoms, into a smaller set of CG macrostates. In order to proceed to the clustering, we first need a notion of distance between a pair of coarse-grained structures, which is here provided by the CG RMSD:
\small
\begin{equation}\label{eq:cg_rmsd}
    \text{RMSD}^{\text{CG}} ({\bf M}({\bf x}) , {\bf M}({\bf y})) = \sqrt{\frac{1}{N} \sum_{I=1}^{N} ({\bf M}_I ({\bf x}) - \mathcal{RT} {\bf M}_I ({\bf y}))^{2}}
\end{equation}
\normalsize
where ${\bf x}$ and ${\bf y}$ are fully atomistic configurations and $\mathcal{RT}$ is the optimal rigid roto-translation that superimposes the two mapped structures. Setting the value of parameter \texttt{rsd} to $1$, the unweighted version of $\text{RMSD}^{\text{CG}}$ is employed as a similarity measure:
\small
\begin{equation}\label{eq:rsd}
 \text{RSD}^{\text{CG}} ({\bf M}({\bf x}) , {\bf M}({\bf y})) = \sqrt{\sum_{I=1}^{N} \left( {\bf M}_I ({\bf x}) - \mathcal{RT} {\bf M}_I ({\bf y}) \right)^{2}}.
\end{equation}
\normalsize

Once the calculation of $ \text{RMSD}^{\text{CG}}$ (or $\text{RSD}^{\text{CG}}$) is carried out for each pair of structures that must be compared, we have a full distance matrix over which a clustering algorithm is applied. When the number of pairs of structures to be compared exceeds the hundreds of thousands, the calculation of the $\text{RMSD}^{\text{CG}}$ distance matrix necessarily slows down due to the huge number of alignments to be performed to superimpose each structure onto each other. This slowdown is particularly critical for the subprograms \emph{\textbf{optimize}} and \emph{\textbf{optimize\_kl}}, in which the calculation of such matrix has to be iterated for thousands of \texttt{MC\_steps}. In Ref.~\citenum{giulini2020information} some of us proposed an approximation that allows one to partially circumvent this problem: in the case of large biological molecules, it is reasonable to assume that the optimal alignment $\mathcal{RT}$ of two CG conformations does not change much if these differ by one or few retained atoms. Therefore, one can keep the alignment constant for a number (\texttt{rotmats\_period}, see Tab.~\ref{tab:params}) of \texttt{MC\_steps}, substantially speeding-up the calculation of the $\text{RMSD}^{\text{CG}}$ distance matrix at the cost of a minimal and controllable error \cite{giulini2020information}.

As for the clustering algorithm, we employ average linkage, agglomerative hierarchical clustering (UPGMA \cite{sokal1958statistical}) to create a dendrogram: at the lowest level of the hierarchy, we have a CG macrostate for each mapped structure, while at the top level there is only one ${\bf R}$ containing all the available structures. Therefore, a \texttt{criterion} (see Tab.~\ref{tab:params}) is required to cut this dendrogram and map each microstate to the corresponding coarse-grained configurational cluster. \texttt{criterion} can assume four values, each one associated with a slightly different choice for cutting the dendrogram.

\begin{figure*}
\centering
                \includegraphics[width=.9\textwidth]{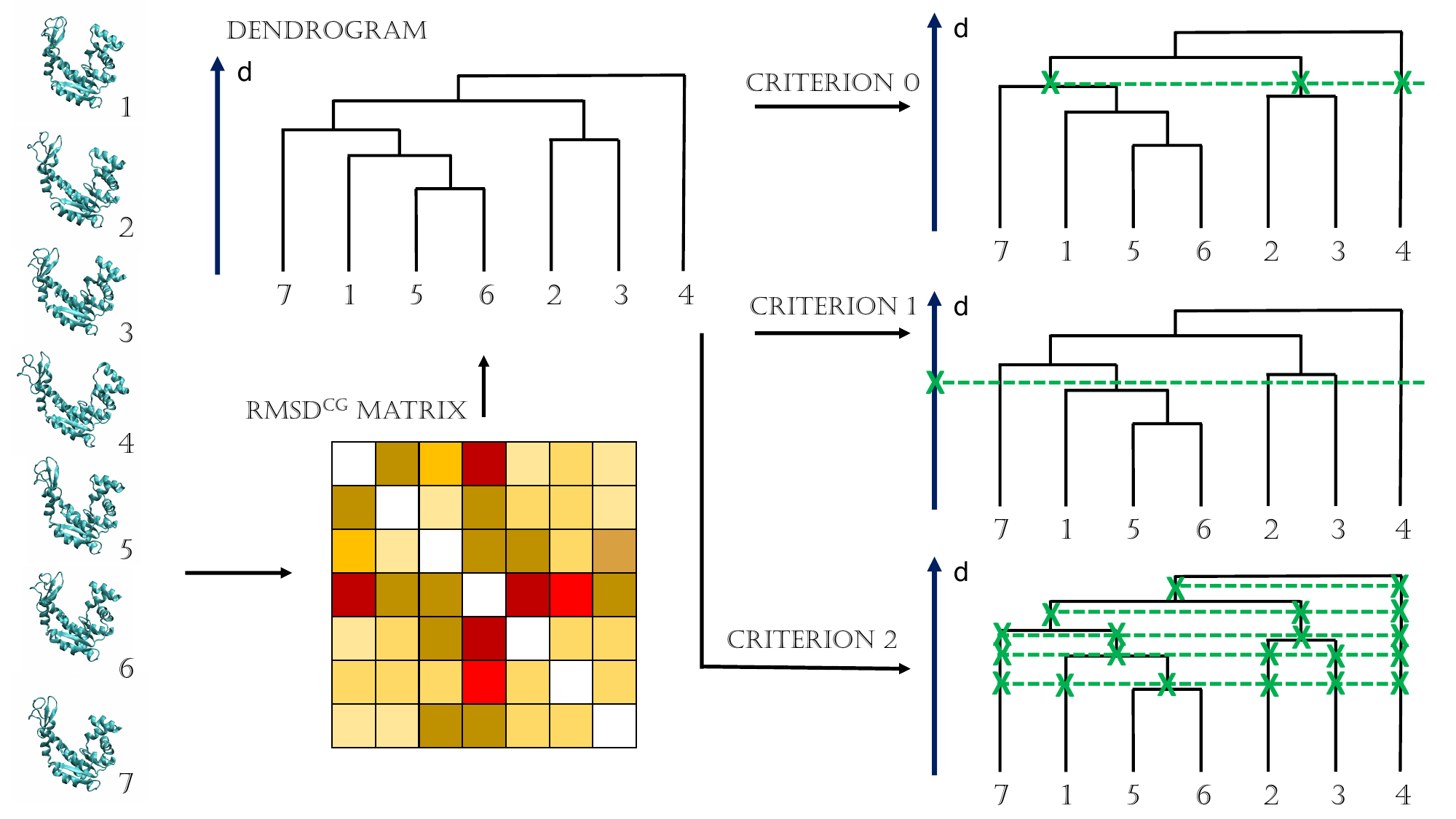}
                \caption{\label{fig:crit0-2}Schematic representation of criteria $0$, $1$, and $2$ for conformational clustering. These are equivalent in the first stage of the procedure, where a $\text{RMSD}^{\text{CG}}$ matrix is calculated between all the configurations (\texttt{frames}, see Tab.~\ref{tab:params}) of a full-atom MD trajectory, observed through the glasses of a CG mapping. From this typically large matrix, the full dendrogram is constructed using the average linkage prescription. Then, conformational clusters can be selected in three manners, namely ($0$) cutting the dendrogram when \texttt{nclust} (equal to $3$ in this case) leaves are present; $1$) cutting the dendrogram when a certain value of cophenetic \texttt{distance} (on the ordinate) is reached, irrespectively of the number of leaves; $2$) applying the procedure ($0$( for a set of $5$ evenly spaced values of the number of clusters ($\{2,3,4,5,6\}$ in this case), determined by parameters \texttt{min\_nclust} and \texttt{max\_nclust} ($2$ and $6$ in this figure).}
\end{figure*}

\begin{figure*}
\centering
                \includegraphics[width=.9\textwidth]{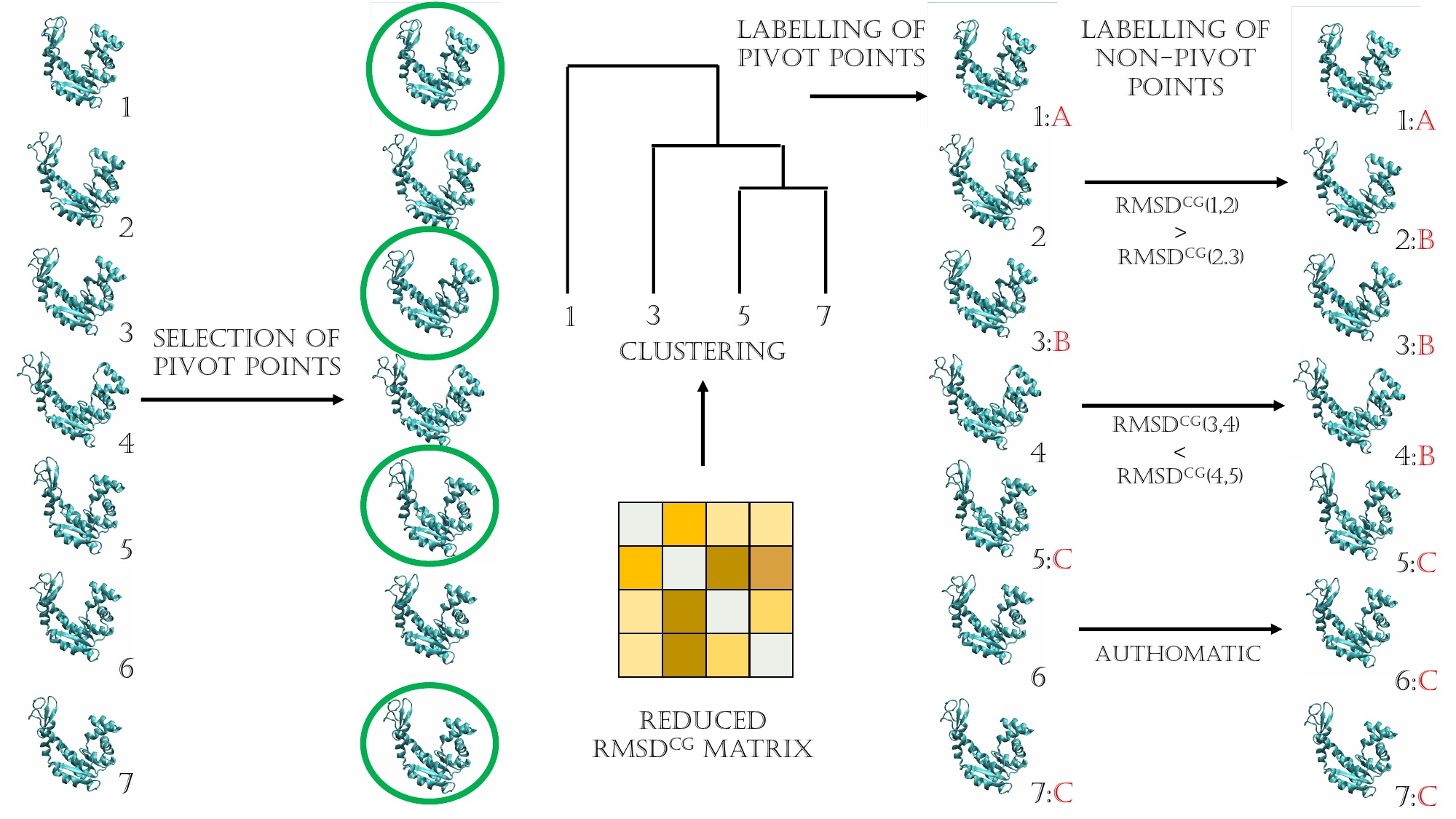}
                \caption{\label{fig:crit3}Graphical description of \texttt{criterion} $3$ for an accelerated clustering of the conformational space. The \texttt{stride} parameter (Tab.~\ref{tab:params}) is equal to $2$ in this case, meaning that $4$ pivot points are considered. The reduced $\text{RMSD}^{\text{CG}}$ matrix and dendrogram are computed taking into account only the coordinates of the selected conformations. Upon clustering, labels of the non-pivot points are assigned based on their proximity with respect to the two closest pivots. If the latter share the same label, as it is for configurations $5$ and $7$ in this example, the intermediate structures are automatically labelled.}
\end{figure*}

\begin{itemize}[leftmargin=0pt]

\item $\text{\texttt{criterion}} = 0$ Analogously to the \textit{maxclust} criterion in \textit{scipy}, a fixed number of coarse-grained macrostates is retrieved. The dendrogram is cut when the number of clusters matches the input parameter \texttt{nclust} (Tab.~\ref{tab:params});

\item $\text{\texttt{criterion}} = 1$ Corresponding to the \textit{distance} criterion in \textit{scipy}, the number of coarse-grained macrostates is not fixed, but rather determined by the cophenetic distance. More specifically, the algorithm cuts the dendrogram when MD configurations in each cluster possess a cophenetic distance lower than the input parameter \texttt{distance} (Tab.~\ref{tab:params}). This choice can be effectively employed in order to observe the scaling of $S_{map}$ with the number of CG sites. In the latter context the \texttt{rsd} parameter must be set to $1$ to make use of the unweighted RMSD as a similarity measure between CG structures;

\item $\text{\texttt{criterion}} = 2$ The iteration of $\text{\texttt{criterion}} = 0$ for five integers between input parameters \texttt{min\_nclust} and \texttt{max\_nclust} (Tab.~\ref{tab:params}). This prescription is used to compute \asmap{}
\begin{equation}
\label{eq:ave_smap}
\asmap{}({\bf M}) = \frac{1}{5} \sum_{K \in \mathcal{K}} {S}_{map}^K({\bf M})
\end{equation}
as in Refs.~\cite{giulini2020information} and \cite{errica2021deep}, with the purpose of increasing the robustness of the SA procedure employed in the mapping optimization. Here, $\mathcal{K}$ is the set of integers employed and ${S}_{map}^K$ is the mapping entropy associated with a specific choice of the number of clusters $K$.

A pictorial representation of criteria $0$, $1$, and $2$ is sketched in Fig. \ref{fig:crit0-2}.

\item $\text{\texttt{criterion}} = 3$ A fast version of $\text{\texttt{criterion}} = 0$ that can be used only when a continuous trajectory of the system is provided in input. In this case, the algorithm computes the pairwise $\text{RMSD}^{\text{CG}}$ matrix between a subset of the overall configurations of the trajectory, that is, one every \texttt{stride} (Tab.~\ref{tab:params}) configurations. For example, if $\text{\texttt{frames}} = 101$ and $\text{\texttt{stride}} = 50$, only \quotes{pivot} configurations number $1$, $51$ and $101$ are considered in the pairwise alignments. Subsequently, standard hierarchical clustering applied to this reduced matrix assigns the coarse-grained macrostate to each \textit{pivot} configuration. Then, the remaining data points are labelled using a simple prescription: if the previous and following pivot configurations possess the same label, the latter is assigned to all the intermediate structures. Instead, if the two pivot points have been labelled differently by the clustering algorithm, each intermediate structure is assigned to the same cluster of the closest pivot, that is, the one corresponding to the lowest $\text{RMSD}^{\text{CG}}$. This approximation guarantees a substantial speed-up to the overall calculation, as the computation of the $\text{RMSD}^{\text{CG}}$ matrix and the following clustering are the most cumbersome tasks, scaling quadratically with the number of frames of the trajectory. More specifically, given a certain value of \texttt{frames}, $f$, and \texttt{stride}, $s$, the overall number of pairwise alignments, $N_a$, in the worst case scenario is given by:
\begin{equation}
N_{a} = \frac{N_p (N_p -1)}{2} + 2(f-N_p)
\end{equation}
where $N_p = \frac{f}{s} + 1$ is the total number of pivot points. As for the clustering procedure, its high computational cost ($\mathcal{O}(f^2 log f))$ makes this criterion extremely appealing. As an example, $s=10$ corresponds to a speed-up factor approximately equal to $300$. This procedure is schematically illustrated in Fig.~\ref{fig:crit3}, where the computational gain arising by employing this criterion is made evident by the shrinkage of both $\text{RMSD}^{\text{CG}}$ matrix and dendrogram.

\end{itemize}

\begin{acknowledgement}
The authors are indebted with M. Mele for an attentive reading of the manuscript.
RP acknowledges support from ICSC - Centro Nazionale di Ricerca in HPC, Big Data and Quantum Computing, funded by the European Union under NextGenerationEU. Views and opinions expressed are however those of the author(s) only and do not necessarily reflect those of the European Union or The European Research Executive Agency. Neither the European Union nor the granting authority can be held responsible for them.
\end{acknowledgement}

\section*{Author contributions}
RP proposed the study; RM, RP, MG conceived the work plan and proposed the method; MG and RM implemented the software and performed the test; RF and LT contributed to the software development. All authors contributed to the analysis and interpretation of the data. All authors drafted the paper, reviewed the results, and approved the final version of the manuscript.






\section*{Data and software availability}
Excogito is available for download at \href{https://github.com/potestiolab/excogito}{https://github.com/potestiolab/excogito}, including the manual and tutorials.




\bibliography{main}

\providecommand{\latin}[1]{#1}
\makeatletter
\providecommand{\doi}
  {\begingroup\let\do\@makeother\dospecials
  \catcode`\{=1 \catcode`\}=2 \doi@aux}
\providecommand{\doi@aux}[1]{\endgroup\texttt{#1}}
\makeatother
\providecommand*\mcitethebibliography{\thebibliography}
\csname @ifundefined\endcsname{endmcitethebibliography}
  {\let\endmcitethebibliography\endthebibliography}{}
\begin{mcitethebibliography}{45}
\providecommand*\natexlab[1]{#1}
\providecommand*\mciteSetBstSublistMode[1]{}
\providecommand*\mciteSetBstMaxWidthForm[2]{}
\providecommand*\mciteBstWouldAddEndPuncttrue
  {\def\EndOfBibitem{\unskip.}}
\providecommand*\mciteBstWouldAddEndPunctfalse
  {\let\EndOfBibitem\relax}
\providecommand*\mciteSetBstMidEndSepPunct[3]{}
\providecommand*\mciteSetBstSublistLabelBeginEnd[3]{}
\providecommand*\EndOfBibitem{}
\mciteSetBstSublistMode{f}
\mciteSetBstMaxWidthForm{subitem}{(\alph{mcitesubitemcount})}
\mciteSetBstSublistLabelBeginEnd
  {\mcitemaxwidthsubitemform\space}
  {\relax}
  {\relax}

\bibitem[Noid(2013)]{noid2013perspective}
Noid,~W.~G. Perspective: Coarse-grained models for biomolecular systems.
  \emph{The Journal of chemical physics} \textbf{2013}, \emph{139},
  09B201\_1\relax
\mciteBstWouldAddEndPuncttrue
\mciteSetBstMidEndSepPunct{\mcitedefaultmidpunct}
{\mcitedefaultendpunct}{\mcitedefaultseppunct}\relax
\EndOfBibitem
\bibitem[Kmiecik \latin{et~al.}(2016)Kmiecik, Gront, Kolinski, Wieteska, Dawid,
  and Kolinski]{kmiecik2016coarse}
Kmiecik,~S.; Gront,~D.; Kolinski,~M.; Wieteska,~L.; Dawid,~A.~E.; Kolinski,~A.
  Coarse-grained protein models and their applications. \emph{Chemical reviews}
  \textbf{2016}, \emph{116}, 7898--7936\relax
\mciteBstWouldAddEndPuncttrue
\mciteSetBstMidEndSepPunct{\mcitedefaultmidpunct}
{\mcitedefaultendpunct}{\mcitedefaultseppunct}\relax
\EndOfBibitem
\bibitem[Giulini \latin{et~al.}(2021)Giulini, Rigoli, Mattiotti, Menichetti,
  Tarenzi, Fiorentini, and Potestio]{giulini2021system}
Giulini,~M.; Rigoli,~M.; Mattiotti,~G.; Menichetti,~R.; Tarenzi,~T.;
  Fiorentini,~R.; Potestio,~R. From System Modeling to System Analysis: The
  Impact of Resolution Level and Resolution Distribution in the Computer-Aided
  Investigation of Biomolecules. \emph{Frontiers in Molecular Biosciences}
  \textbf{2021}, \emph{8}\relax
\mciteBstWouldAddEndPuncttrue
\mciteSetBstMidEndSepPunct{\mcitedefaultmidpunct}
{\mcitedefaultendpunct}{\mcitedefaultseppunct}\relax
\EndOfBibitem
\bibitem[Dhamankar and Webb(2021)Dhamankar, and Webb]{dhamankar2021chemically}
Dhamankar,~S.; Webb,~M.~A. Chemically specific coarse-graining of polymers:
  methods and prospects. \emph{Journal of Polymer Science} \textbf{2021},
  \emph{59}, 2613--2643\relax
\mciteBstWouldAddEndPuncttrue
\mciteSetBstMidEndSepPunct{\mcitedefaultmidpunct}
{\mcitedefaultendpunct}{\mcitedefaultseppunct}\relax
\EndOfBibitem
\bibitem[Noid(2023)]{noid2023perspective}
Noid,~W. Perspective: Advances, challenges, and insight for predictive
  coarse-grained models. \emph{The Journal of Physical Chemistry B}
  \textbf{2023}, \relax
\mciteBstWouldAddEndPunctfalse
\mciteSetBstMidEndSepPunct{\mcitedefaultmidpunct}
{}{\mcitedefaultseppunct}\relax
\EndOfBibitem
\bibitem[Marrink \latin{et~al.}(2007)Marrink, Risselada, Yefimov, Tieleman, and
  De~Vries]{marrink2007martini}
Marrink,~S.~J.; Risselada,~H.~J.; Yefimov,~S.; Tieleman,~D.~P.; De~Vries,~A.~H.
  The MARTINI force field: coarse grained model for biomolecular simulations.
  \emph{The journal of physical chemistry B} \textbf{2007}, \emph{111},
  7812--7824\relax
\mciteBstWouldAddEndPuncttrue
\mciteSetBstMidEndSepPunct{\mcitedefaultmidpunct}
{\mcitedefaultendpunct}{\mcitedefaultseppunct}\relax
\EndOfBibitem
\bibitem[Wu \latin{et~al.}(2010)Wu, Cui, and Yethiraj]{wu2010new}
Wu,~Z.; Cui,~Q.; Yethiraj,~A. A new coarse-grained model for water: the
  importance of electrostatic interactions. \emph{The Journal of Physical
  Chemistry B} \textbf{2010}, \emph{114}, 10524--10529\relax
\mciteBstWouldAddEndPuncttrue
\mciteSetBstMidEndSepPunct{\mcitedefaultmidpunct}
{\mcitedefaultendpunct}{\mcitedefaultseppunct}\relax
\EndOfBibitem
\bibitem[Hadley and McCabe(2012)Hadley, and McCabe]{hadley2012coarse}
Hadley,~K.~R.; McCabe,~C. Coarse-grained molecular models of water: a review.
  \emph{Molecular simulation} \textbf{2012}, \emph{38}, 671--681\relax
\mciteBstWouldAddEndPuncttrue
\mciteSetBstMidEndSepPunct{\mcitedefaultmidpunct}
{\mcitedefaultendpunct}{\mcitedefaultseppunct}\relax
\EndOfBibitem
\bibitem[Ouldridge \latin{et~al.}(2010)Ouldridge, Louis, and
  Doye]{ouldridge2010dna}
Ouldridge,~T.~E.; Louis,~A.~A.; Doye,~J.~P. DNA nanotweezers studied with a
  coarse-grained model of DNA. \emph{Physical Review Letters} \textbf{2010},
  \emph{104}, 178101\relax
\mciteBstWouldAddEndPuncttrue
\mciteSetBstMidEndSepPunct{\mcitedefaultmidpunct}
{\mcitedefaultendpunct}{\mcitedefaultseppunct}\relax
\EndOfBibitem
\bibitem[Marrink \latin{et~al.}(2023)Marrink, Monticelli, Melo, Alessandri,
  Tieleman, and Souza]{marrink2023two}
Marrink,~S.~J.; Monticelli,~L.; Melo,~M.~N.; Alessandri,~R.; Tieleman,~D.~P.;
  Souza,~P.~C. Two decades of Martini: Better beads, broader scope. \emph{Wiley
  Interdisciplinary Reviews: Computational Molecular Science} \textbf{2023},
  \emph{13}, e1620\relax
\mciteBstWouldAddEndPuncttrue
\mciteSetBstMidEndSepPunct{\mcitedefaultmidpunct}
{\mcitedefaultendpunct}{\mcitedefaultseppunct}\relax
\EndOfBibitem
\bibitem[Earnest \latin{et~al.}(2018)Earnest, Cole, and
  Luthey-Schulten]{earnest2018simulating}
Earnest,~T.~M.; Cole,~J.~A.; Luthey-Schulten,~Z. Simulating biological
  processes: stochastic physics from whole cells to colonies. \emph{Reports on
  Progress in Physics} \textbf{2018}, \emph{81}, 052601\relax
\mciteBstWouldAddEndPuncttrue
\mciteSetBstMidEndSepPunct{\mcitedefaultmidpunct}
{\mcitedefaultendpunct}{\mcitedefaultseppunct}\relax
\EndOfBibitem
\bibitem[Thornburg \latin{et~al.}(2022)Thornburg, Bianchi, Brier, Gilbert,
  Earnest, Melo, Safronova, S{\'a}enz, Cook, Wise, Hutchison, Smith, Glass, and
  Luthey-Schulten]{thornburg2022fundamental}
Thornburg,~Z.~R.; Bianchi,~D.~M.; Brier,~T.~A.; Gilbert,~B.~R.; Earnest,~T.~M.;
  Melo,~M.~C.; Safronova,~N.; S{\'a}enz,~J.~P.; Cook,~A.~T.; Wise,~K.~S.;
  Hutchison,~C. A.~I.; Smith,~H.~O.; Glass,~J.~I.; Luthey-Schulten,~Z.
  Fundamental behaviors emerge from simulations of a living minimal cell.
  \emph{Cell} \textbf{2022}, \emph{185}, 345--360\relax
\mciteBstWouldAddEndPuncttrue
\mciteSetBstMidEndSepPunct{\mcitedefaultmidpunct}
{\mcitedefaultendpunct}{\mcitedefaultseppunct}\relax
\EndOfBibitem
\bibitem[Luthey-Schulten \latin{et~al.}(2022)Luthey-Schulten, Thornburg, and
  Gilbert]{luthey2022integrating}
Luthey-Schulten,~Z.; Thornburg,~Z.~R.; Gilbert,~B.~R. Integrating cellular and
  molecular structures and dynamics into whole-cell models. \emph{Current
  Opinion in Structural Biology} \textbf{2022}, \emph{75}, 102392\relax
\mciteBstWouldAddEndPuncttrue
\mciteSetBstMidEndSepPunct{\mcitedefaultmidpunct}
{\mcitedefaultendpunct}{\mcitedefaultseppunct}\relax
\EndOfBibitem
\bibitem[Stevens \latin{et~al.}(2023)Stevens, Gr{\"u}newald, van Tilburg,
  K{\"o}nig, Gilbert, Brier, Thornburg, Luthey-Schulten, and
  Marrink]{stevens2023molecular}
Stevens,~J.~A.; Gr{\"u}newald,~F.; van Tilburg,~P.~M.; K{\"o}nig,~M.;
  Gilbert,~B.~R.; Brier,~T.~A.; Thornburg,~Z.~R.; Luthey-Schulten,~Z.;
  Marrink,~S.~J. Molecular dynamics simulation of an entire cell.
  \emph{Frontiers in Chemistry} \textbf{2023}, \emph{11}, 1106495\relax
\mciteBstWouldAddEndPuncttrue
\mciteSetBstMidEndSepPunct{\mcitedefaultmidpunct}
{\mcitedefaultendpunct}{\mcitedefaultseppunct}\relax
\EndOfBibitem
\bibitem[Potestio \latin{et~al.}(2014)Potestio, Peter, and
  Kremer]{potestio2014computer}
Potestio,~R.; Peter,~C.; Kremer,~K. Computer simulations of soft matter:
  Linking the scales. \emph{Entropy} \textbf{2014}, \emph{16}, 4199--4245\relax
\mciteBstWouldAddEndPuncttrue
\mciteSetBstMidEndSepPunct{\mcitedefaultmidpunct}
{\mcitedefaultendpunct}{\mcitedefaultseppunct}\relax
\EndOfBibitem
\bibitem[Wilson(1971)]{wilson1971renormalization}
Wilson,~K.~G. Renormalization group and critical phenomena. I. Renormalization
  group and the Kadanoff scaling picture. \emph{Physical review B}
  \textbf{1971}, \emph{4}, 3174\relax
\mciteBstWouldAddEndPuncttrue
\mciteSetBstMidEndSepPunct{\mcitedefaultmidpunct}
{\mcitedefaultendpunct}{\mcitedefaultseppunct}\relax
\EndOfBibitem
\bibitem[Kadanoff(1990)]{kadanoff1990scaling}
Kadanoff,~L.~P. Scaling and universality in statistical physics. \emph{Physica
  A: Statistical Mechanics and its Applications} \textbf{1990}, \emph{163},
  1--14\relax
\mciteBstWouldAddEndPuncttrue
\mciteSetBstMidEndSepPunct{\mcitedefaultmidpunct}
{\mcitedefaultendpunct}{\mcitedefaultseppunct}\relax
\EndOfBibitem
\bibitem[Efrati \latin{et~al.}(2014)Efrati, Wang, Kolan, and
  Kadanoff]{efrati2014real}
Efrati,~E.; Wang,~Z.; Kolan,~A.; Kadanoff,~L.~P. Real-space renormalization in
  statistical mechanics. \emph{Reviews of Modern Physics} \textbf{2014},
  \emph{86}, 647\relax
\mciteBstWouldAddEndPuncttrue
\mciteSetBstMidEndSepPunct{\mcitedefaultmidpunct}
{\mcitedefaultendpunct}{\mcitedefaultseppunct}\relax
\EndOfBibitem
\bibitem[Adami(1995)]{adami1995self}
Adami,~C. Self-organized criticality in living systems. \emph{Physics Letters
  A} \textbf{1995}, \emph{203}, 29--32\relax
\mciteBstWouldAddEndPuncttrue
\mciteSetBstMidEndSepPunct{\mcitedefaultmidpunct}
{\mcitedefaultendpunct}{\mcitedefaultseppunct}\relax
\EndOfBibitem
\bibitem[Mora and Bialek(2011)Mora, and Bialek]{mora2011biological}
Mora,~T.; Bialek,~W. Are biological systems poised at criticality?
  \emph{Journal of Statistical Physics} \textbf{2011}, \emph{144},
  268--302\relax
\mciteBstWouldAddEndPuncttrue
\mciteSetBstMidEndSepPunct{\mcitedefaultmidpunct}
{\mcitedefaultendpunct}{\mcitedefaultseppunct}\relax
\EndOfBibitem
\bibitem[Marsili(2020)]{marsili2020importance}
Marsili,~M. On the importance of being critical. \emph{Europhysics News}
  \textbf{2020}, \emph{51}, 42--44\relax
\mciteBstWouldAddEndPuncttrue
\mciteSetBstMidEndSepPunct{\mcitedefaultmidpunct}
{\mcitedefaultendpunct}{\mcitedefaultseppunct}\relax
\EndOfBibitem
\bibitem[Jin \latin{et~al.}(2022)Jin, Pak, Durumeric, Loose, and
  Voth]{jin2022bottom}
Jin,~J.; Pak,~A.~J.; Durumeric,~A.~E.; Loose,~T.~D.; Voth,~G.~A. Bottom-up
  coarse-graining: Principles and perspectives. \emph{Journal of Chemical
  Theory and Computation} \textbf{2022}, \emph{18}, 5759--5791\relax
\mciteBstWouldAddEndPuncttrue
\mciteSetBstMidEndSepPunct{\mcitedefaultmidpunct}
{\mcitedefaultendpunct}{\mcitedefaultseppunct}\relax
\EndOfBibitem
\bibitem[Brini \latin{et~al.}(2013)Brini, Algaer, Ganguly, Li,
  Rodr{\'\i}guez-Ropero, and van~der Vegt]{brini2013systematic}
Brini,~E.; Algaer,~E.~A.; Ganguly,~P.; Li,~C.; Rodr{\'\i}guez-Ropero,~F.;
  van~der Vegt,~N.~F. Systematic coarse-graining methods for soft matter
  simulations--a review. \emph{Soft Matter} \textbf{2013}, \emph{9},
  2108--2119\relax
\mciteBstWouldAddEndPuncttrue
\mciteSetBstMidEndSepPunct{\mcitedefaultmidpunct}
{\mcitedefaultendpunct}{\mcitedefaultseppunct}\relax
\EndOfBibitem
\bibitem[Saunders and Voth(2013)Saunders, and Voth]{saunders2013coarse}
Saunders,~M.~G.; Voth,~G.~A. Coarse-graining methods for computational biology.
  \emph{Annual review of biophysics} \textbf{2013}, \emph{42}, 73--93\relax
\mciteBstWouldAddEndPuncttrue
\mciteSetBstMidEndSepPunct{\mcitedefaultmidpunct}
{\mcitedefaultendpunct}{\mcitedefaultseppunct}\relax
\EndOfBibitem
\bibitem[Ing{\'o}lfsson \latin{et~al.}(2014)Ing{\'o}lfsson, Lopez, Uusitalo,
  de~Jong, Gopal, Periole, and Marrink]{ingolfsson2014power}
Ing{\'o}lfsson,~H.~I.; Lopez,~C.~A.; Uusitalo,~J.~J.; de~Jong,~D.~H.;
  Gopal,~S.~M.; Periole,~X.; Marrink,~S.~J. The power of coarse graining in
  biomolecular simulations. \emph{Wiley Interdisciplinary Reviews:
  Computational Molecular Science} \textbf{2014}, \emph{4}, 225--248\relax
\mciteBstWouldAddEndPuncttrue
\mciteSetBstMidEndSepPunct{\mcitedefaultmidpunct}
{\mcitedefaultendpunct}{\mcitedefaultseppunct}\relax
\EndOfBibitem
\bibitem[Rudzinski and Noid(2014)Rudzinski, and
  Noid]{rudzinski2014investigation}
Rudzinski,~J.~F.; Noid,~W.~G. Investigation of coarse-grained mappings via an
  iterative generalized Yvon--Born--Green method. \emph{The Journal of Physical
  Chemistry B} \textbf{2014}, \emph{118}, 8295--8312\relax
\mciteBstWouldAddEndPuncttrue
\mciteSetBstMidEndSepPunct{\mcitedefaultmidpunct}
{\mcitedefaultendpunct}{\mcitedefaultseppunct}\relax
\EndOfBibitem
\bibitem[Wang and G{\'o}mez-Bombarelli(2019)Wang, and
  G{\'o}mez-Bombarelli]{wang2019coarse}
Wang,~W.; G{\'o}mez-Bombarelli,~R. Coarse-graining auto-encoders for molecular
  dynamics. \emph{npj Computational Materials} \textbf{2019}, \emph{5},
  125\relax
\mciteBstWouldAddEndPuncttrue
\mciteSetBstMidEndSepPunct{\mcitedefaultmidpunct}
{\mcitedefaultendpunct}{\mcitedefaultseppunct}\relax
\EndOfBibitem
\bibitem[Yang \latin{et~al.}(2023)Yang, Templeton, Rosenberger, Bittracher,
  Nuuske, No{\'e}, and Clementi]{yang2023slicing}
Yang,~W.; Templeton,~C.; Rosenberger,~D.; Bittracher,~A.; Nuuske,~F.;
  No{\'e},~F.; Clementi,~C. Slicing and Dicing: Optimal Coarse-Grained
  Representation to Preserve Molecular Kinetics. \emph{ACS Central Science}
  \textbf{2023}, \emph{9}, 186--196\relax
\mciteBstWouldAddEndPuncttrue
\mciteSetBstMidEndSepPunct{\mcitedefaultmidpunct}
{\mcitedefaultendpunct}{\mcitedefaultseppunct}\relax
\EndOfBibitem
\bibitem[Shell(2008)]{shell2008relative}
Shell,~M.~S. The relative entropy is fundamental to multiscale and inverse
  thermodynamic problems. \emph{The Journal of chemical physics} \textbf{2008},
  \emph{129}, 144108\relax
\mciteBstWouldAddEndPuncttrue
\mciteSetBstMidEndSepPunct{\mcitedefaultmidpunct}
{\mcitedefaultendpunct}{\mcitedefaultseppunct}\relax
\EndOfBibitem
\bibitem[Foley \latin{et~al.}(2015)Foley, Shell, and Noid]{foley2015impact}
Foley,~T.~T.; Shell,~M.~S.; Noid,~W.~G. The impact of resolution upon entropy
  and information in coarse-grained models. \emph{The Journal of chemical
  physics} \textbf{2015}, \emph{143}, 12B601\_1\relax
\mciteBstWouldAddEndPuncttrue
\mciteSetBstMidEndSepPunct{\mcitedefaultmidpunct}
{\mcitedefaultendpunct}{\mcitedefaultseppunct}\relax
\EndOfBibitem
\bibitem[Giulini \latin{et~al.}(2020)Giulini, Menichetti, Shell, and
  Potestio]{giulini2020information}
Giulini,~M.; Menichetti,~R.; Shell,~M.~S.; Potestio,~R. An
  Information-Theory-Based Approach for Optimal Model Reduction of
  Biomolecules. \emph{Journal of chemical theory and computation}
  \textbf{2020}, \emph{16}, 6795--6813\relax
\mciteBstWouldAddEndPuncttrue
\mciteSetBstMidEndSepPunct{\mcitedefaultmidpunct}
{\mcitedefaultendpunct}{\mcitedefaultseppunct}\relax
\EndOfBibitem
\bibitem[Kidder \latin{et~al.}(2021)Kidder, Szukalo, and
  Noid]{kidder2021energetic}
Kidder,~K.~M.; Szukalo,~R.~J.; Noid,~W. Energetic and entropic considerations
  for coarse-graining. \emph{The European Physical Journal B} \textbf{2021},
  \emph{94}, 153\relax
\mciteBstWouldAddEndPuncttrue
\mciteSetBstMidEndSepPunct{\mcitedefaultmidpunct}
{\mcitedefaultendpunct}{\mcitedefaultseppunct}\relax
\EndOfBibitem
\bibitem[Holtzman \latin{et~al.}(2022)Holtzman, Giulini, and
  Potestio]{holtzman2022making}
Holtzman,~R.; Giulini,~M.; Potestio,~R. Making sense of complex systems through
  resolution, relevance, and mapping entropy. \emph{Physical Review E}
  \textbf{2022}, \emph{106}, 044101\relax
\mciteBstWouldAddEndPuncttrue
\mciteSetBstMidEndSepPunct{\mcitedefaultmidpunct}
{\mcitedefaultendpunct}{\mcitedefaultseppunct}\relax
\EndOfBibitem
\bibitem[Menichetti \latin{et~al.}(2021)Menichetti, Giulini, and
  Potestio]{menichetti2021journey}
Menichetti,~R.; Giulini,~M.; Potestio,~R. A journey through mapping space:
  characterising the statistical and metric properties of reduced
  representations of macromolecules. \emph{The European Physical Journal B}
  \textbf{2021}, \emph{94}, 204\relax
\mciteBstWouldAddEndPuncttrue
\mciteSetBstMidEndSepPunct{\mcitedefaultmidpunct}
{\mcitedefaultendpunct}{\mcitedefaultseppunct}\relax
\EndOfBibitem
\bibitem[Foley \latin{et~al.}(2020)Foley, Kidder, Shell, and
  Noid]{foley2020exploring}
Foley,~T.~T.; Kidder,~K.~M.; Shell,~M.~S.; Noid,~W. Exploring the landscape of
  model representations. \emph{Proceedings of the National Academy of Sciences}
  \textbf{2020}, \emph{117}, 24061--24068\relax
\mciteBstWouldAddEndPuncttrue
\mciteSetBstMidEndSepPunct{\mcitedefaultmidpunct}
{\mcitedefaultendpunct}{\mcitedefaultseppunct}\relax
\EndOfBibitem
\bibitem[Kidder \latin{et~al.}(2024)Kidder, Shell, and
  Noid]{kidder2024surveying}
Kidder,~K.~M.; Shell,~M.~S.; Noid,~W. Surveying the energy landscape of
  coarse-grained mappings. \emph{The Journal of Chemical Physics}
  \textbf{2024}, \emph{160}\relax
\mciteBstWouldAddEndPuncttrue
\mciteSetBstMidEndSepPunct{\mcitedefaultmidpunct}
{\mcitedefaultendpunct}{\mcitedefaultseppunct}\relax
\EndOfBibitem
\bibitem[Kullback and Leibler(1951)Kullback, and
  Leibler]{kullback1951information}
Kullback,~S.; Leibler,~R.~A. On information and sufficiency. \emph{The annals
  of mathematical statistics} \textbf{1951}, \emph{22}, 79--86\relax
\mciteBstWouldAddEndPuncttrue
\mciteSetBstMidEndSepPunct{\mcitedefaultmidpunct}
{\mcitedefaultendpunct}{\mcitedefaultseppunct}\relax
\EndOfBibitem
\bibitem[Rudzinski and Noid(2011)Rudzinski, and Noid]{rudzinski2011coarse}
Rudzinski,~J.~F.; Noid,~W. Coarse-graining entropy, forces, and structures.
  \emph{The Journal of chemical physics} \textbf{2011}, \emph{135},
  214101\relax
\mciteBstWouldAddEndPuncttrue
\mciteSetBstMidEndSepPunct{\mcitedefaultmidpunct}
{\mcitedefaultendpunct}{\mcitedefaultseppunct}\relax
\EndOfBibitem
\bibitem[Giulini \latin{et~al.}(2024)Giulini, Honorato, Rivera, and
  Bonvin]{giulini2024arctic}
Giulini,~M.; Honorato,~R.~V.; Rivera,~J.~L.; Bonvin,~A.~M. ARCTIC-3D: automatic
  retrieval and clustering of interfaces in complexes from 3D structural
  information. \emph{Communications Biology} \textbf{2024}, \emph{7}, 49\relax
\mciteBstWouldAddEndPuncttrue
\mciteSetBstMidEndSepPunct{\mcitedefaultmidpunct}
{\mcitedefaultendpunct}{\mcitedefaultseppunct}\relax
\EndOfBibitem
\bibitem[Van Der~Spoel \latin{et~al.}(2005)Van Der~Spoel, Lindahl, Hess,
  Groenhof, Mark, and Berendsen]{van2005gromacs}
Van Der~Spoel,~D.; Lindahl,~E.; Hess,~B.; Groenhof,~G.; Mark,~A.~E.;
  Berendsen,~H.~J. GROMACS: fast, flexible, and free. \emph{Journal of
  computational chemistry} \textbf{2005}, \emph{26}, 1701--1718\relax
\mciteBstWouldAddEndPuncttrue
\mciteSetBstMidEndSepPunct{\mcitedefaultmidpunct}
{\mcitedefaultendpunct}{\mcitedefaultseppunct}\relax
\EndOfBibitem
\bibitem[Lindorff-Larsen \latin{et~al.}(2010)Lindorff-Larsen, Piana, Palmo,
  Maragakis, Klepeis, Dror, and Shaw]{lindorff2010improved}
Lindorff-Larsen,~K.; Piana,~S.; Palmo,~K.; Maragakis,~P.; Klepeis,~J.~L.;
  Dror,~R.~O.; Shaw,~D.~E. Improved side-chain torsion potentials for the Amber
  ff99SB protein force field. \emph{Proteins: Structure, Function, and
  Bioinformatics} \textbf{2010}, \emph{78}, 1950--1958\relax
\mciteBstWouldAddEndPuncttrue
\mciteSetBstMidEndSepPunct{\mcitedefaultmidpunct}
{\mcitedefaultendpunct}{\mcitedefaultseppunct}\relax
\EndOfBibitem
\bibitem[Diggins~IV \latin{et~al.}(2018)Diggins~IV, Liu, Deserno, and
  Potestio]{diggins2018optimal}
Diggins~IV,~P.; Liu,~C.; Deserno,~M.; Potestio,~R. Optimal coarse-grained site
  selection in elastic network models of biomolecules. \emph{Journal of
  chemical theory and computation} \textbf{2018}, \emph{15}, 648--664\relax
\mciteBstWouldAddEndPuncttrue
\mciteSetBstMidEndSepPunct{\mcitedefaultmidpunct}
{\mcitedefaultendpunct}{\mcitedefaultseppunct}\relax
\EndOfBibitem
\bibitem[Sokal(1958)]{sokal1958statistical}
Sokal,~R.~R. A statistical method for evaluating systematic relationships.
  \emph{Univ. Kansas, Sci. Bull.} \textbf{1958}, \emph{38}, 1409--1438\relax
\mciteBstWouldAddEndPuncttrue
\mciteSetBstMidEndSepPunct{\mcitedefaultmidpunct}
{\mcitedefaultendpunct}{\mcitedefaultseppunct}\relax
\EndOfBibitem
\bibitem[Errica \latin{et~al.}(2021)Errica, Giulini, Bacciu, Menichetti,
  Micheli, and Potestio]{errica2021deep}
Errica,~F.; Giulini,~M.; Bacciu,~D.; Menichetti,~R.; Micheli,~A.; Potestio,~R.
  A deep graph network-enhanced sampling approach to efficiently explore the
  space of reduced representations of proteins. \emph{Frontiers in Molecular
  Biosciences} \textbf{2021}, \emph{8}, 136\relax
\mciteBstWouldAddEndPuncttrue
\mciteSetBstMidEndSepPunct{\mcitedefaultmidpunct}
{\mcitedefaultendpunct}{\mcitedefaultseppunct}\relax
\EndOfBibitem
\end{mcitethebibliography}



\clearpage

\end{document}